\documentclass[usenatbib]{mn2e}
\usepackage{graphicx}
\usepackage{lscape}
\usepackage{color}
\usepackage{ulem}

\def\lsim{~\rlap{$<$}{\lower 1.0ex\hbox{$\sim$}}}
\def\gsim{~\rlap{$>$}{\lower 1.0ex\hbox{$\sim$}}}

\begin{document}

\title[spin origin in atomic cooling halos]{The origin of spin in galaxies: clues from simulations of atomic cooling halos}

\author[Prieto et al.]{Joaquin Prieto$^{1,5,7}$\thanks{email:joaquin.prieto.brito@gmail.com}, Raul Jimenez$^{2,1,3}$,
Zolt\'{a}n Haiman$^4$, 
Roberto E. Gonz\'{a}lez$^{6,8}$\\
$^1$ ICC, University of Barcelona, Marti i Franques 1, E08028, Barcelona, Spain.\\
$^2$ ICREA.\\
$^3$ Institute for Applied Computational Science, Harvard University, MA 02138, USA.\\
$^4$ Department of Astronomy, Columbia University, 550 West 120th Street, MC 5246, New York, NY 10027, USA.\\
$^5$ Institute of Astronomy and NAO, Bulgarian Academy of Sciences, 72, Tsarigradsko Chaussee Blvd. 1784, Sofia, Bulgaria.\\
$^6$ Instituto de Astrof\'{i}sica, Pontificia Universidad Cat\'{o}lica de Chile, Av. Vicu\~na Mackenna 4860, Santiago, Chile.\\
$^7$ Departamento de Astronom\'{i}a, Universidad de Chile, Casilla 36-D, Santiago, Chile.\\
$^8$ Centro de Astro-Ingenier\'{i}a, Pontificia Universidad Cat\'olica de Chile, Av. Vicu\~na Mackenna 4860, Santiago, Chile}

\maketitle

\begin{abstract}
In order to elucidate the origin of spin in both dark matter and baryons in galaxies, we have performed hydrodynamical 
simulations from cosmological initial conditions. We study atomic cooling haloes in the 
redshift range $100 > z > 9$ with masses of order $10^9{\rm M_{\odot}}$ at redshift $z=10$.  We assume that the gas has 
primordial composition and that ${\rm H_2}$-cooling and prior star-formation in the haloes have been suppressed.  We 
present a comprehensive analysis of the gas and dark matter properties of four halos with very low ($\lambda \approx 0.01$), low 
($\lambda \approx 0.04$), high ($\lambda \approx 0.06$) and very high ($\lambda \approx 0.1$) spin parameter.  
Our main conclusion is that the spin orientation and magnitude is initially well described by tidal torque linear theory, 
but later on is determined by the merging and accretion history of each halo. We provide evidence that the topology of the merging region, i.e. the number of colliding filaments, gives an accurate prediction for the spin of dark matter and gas: halos at the center of knots will have low spin 
while those in the center of filaments will have high spin. The spin of a halo is given 
by $\lambda \approx 0.05 \times \left(\frac{7.6}{\rm number\,\,\, of \,\,\, filaments}\right)^{5.1}$.
\end{abstract}

\begin{keywords}
galaxies: formation --- large-scale structure of the universe --- stars: formation --- turbulence.
\end{keywords}

\section{Introduction}

Hoyle's seminal paper (Hoyle 1949) showed how angular momentum in galaxies can be generated 
via the tidal field of other galaxies. Later, \citet{Doro70} developed the tidal-torque theory within the framework 
of hierarchical galaxy formation, which determines the amplitude and direction of the spin of a dark matter halo based 
on the surrounding dark matter field (see \citet{Schaefer08} for a recent review). Numerical N-body simulations produce 
results that are in good agreement with the theoretical predictions, although linear theory is not always sufficient 
to determine the final angular momentum of a collapsed object (e.g. \citet{BarnesEfstathiou87}; 
\citet{Porciani02a} and references therein). In addition, mergers are expected to 
significantly alter a halo final spin \citep{Vitvitskaetal2002,merger,Davis}.

The amplitude of the dark matter halo spin influences the radius where baryons settle 
into a disc \citep{WhiteRees78,FallEfstathiou80}, thus modifying its density and therefore 
the star formation history of the galaxy. The influence of spin on star formation history has been studied 
in detail \citep{Toomre64,Dalcanton97,JHH97,Mo98,AvilaRese98}.
One interesting feature of tidal-torque theory in hierarchical models \citep{HeavensPeacock88,CatelanPorciani01,Catelanlensing,Crittenden2001,Porciani02a,HahnSpins07}
is the prediction of correlated spin directions and that the spin direction for dark haloes is strongly influenced by the halo 
environment. \citet{PenLeeSeljak2000} reported a detection of galaxy spin correlations at 97\% confidence. Slosar et al. (2009) 
have measured the correlation function of the spin chirality and report, for the first time, a signal at scales $\la 0.5$ 
Mpc $h^{-1}$ at the $2-3 \sigma$ level. \citet{Jimenezspin} showed how star formation in galaxies and current spin are 
correlated, with a $5\sigma$ detection of spin alignment among galaxies which formed most of their stars at $z > 2$. 

Motivated by the theoretical study of \citet{PichonBernardeu1999}, recent interest on the origin of spin in galaxies has focused on 
the effect of baryons on it. In a series of recent simulations \citet{Pichon+2011} and \citet{Codis+2012} have argued that spins may be generated entirely 
in the baryonic component due to the growth of eddies in the turbulence field generated by large-scales ($\ga$ few Mpc) infall. If this was the case, 
it would imply that the final spin of halos is determined by the (random) properties of gas motions and the capability of the gas to generate 
eddies and vorticity. Their study focuses on DM haloes with masses in the range
${\rm 10^{11}M_{\odot}\la M_{vir}\la10^{14}M_{\odot}}$; they found that the low mass haloes tend to have a spin parallel to their associated filament
whereas the high mass haloes tend to have a spin perpendicular to their filament. They explained such correlation based on the shell crossing vorticity 
generation process \citep{PichonBernardeu1999}. In such a picture the low mass haloes are small enough to belong to a vortical region inside a filament. If this is the case, the DM spin (and its galaxy) will be aligned with the vortex vector. On the other hand, when the halo is too big to belong to a single vortex
region its spin will be canceled and the main source of spin will be the angular momentum accreted through filaments. Such process
will tend to create a spin vector perpendicular to the filament.  

\begin{figure*}
\centering
\includegraphics[width=14.0cm,height=22.0cm]{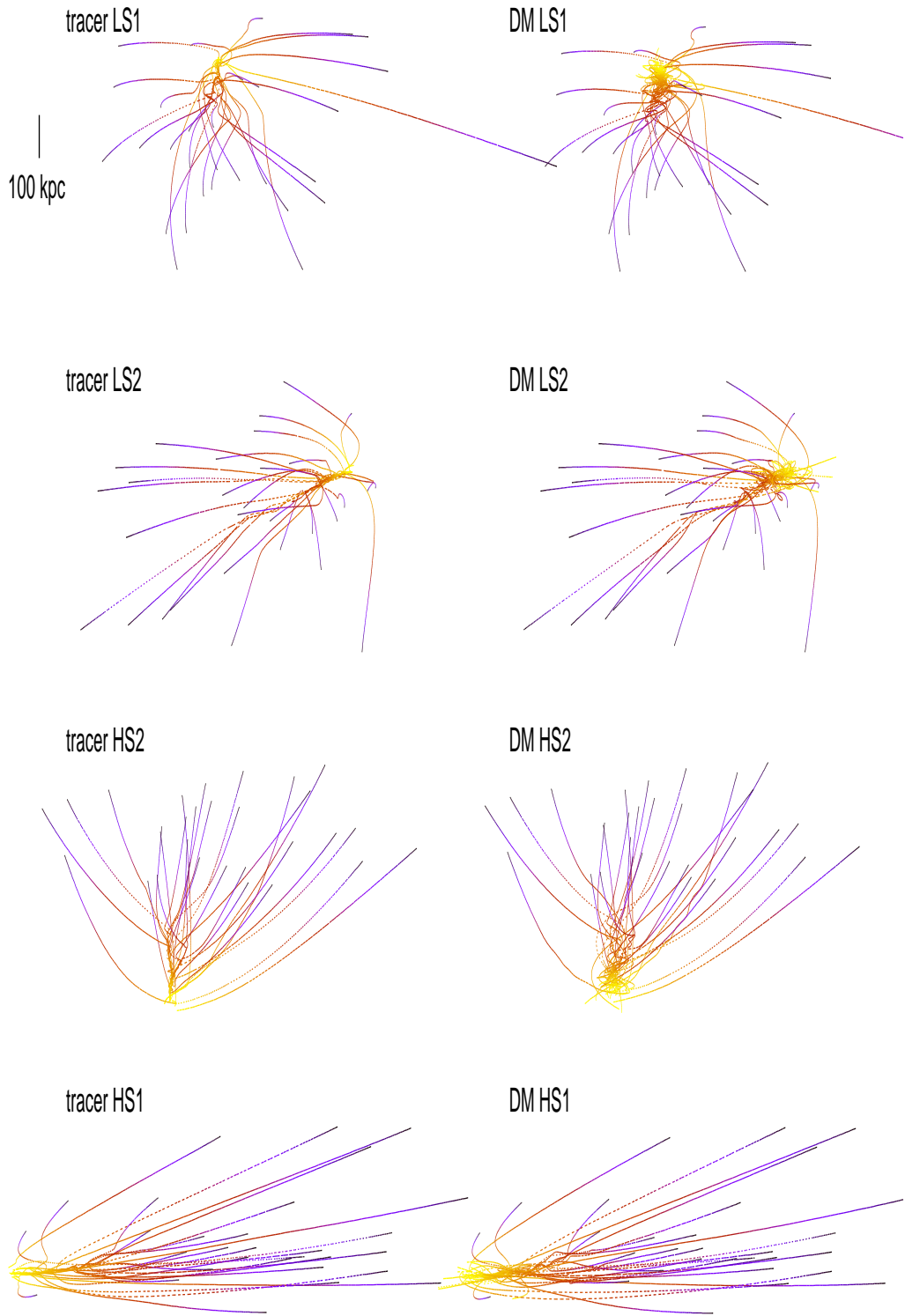}
\caption{Trajectories of gas tracing particles (left) and DM particles (right) for the four different systems 
studied in this work. The trajectories are shown for particles within a radius of $0.5$R$_{\rm vir}$ at the end 
of the simulation. The time evolution is shown from $z\approx100$ until $z\approx9$ and it is color-coded with 
dark meaning early (high redshift) times and light meaning late (low redshift) times. Gas trajectories tend to be 
more concentrated than DM ones. Such behavior is the consequence of the gas
cooling properties.}
\label{fig:traject}
\end{figure*}

\begin{figure*}
\centering
\includegraphics[width=8.2cm,height=9.0cm]{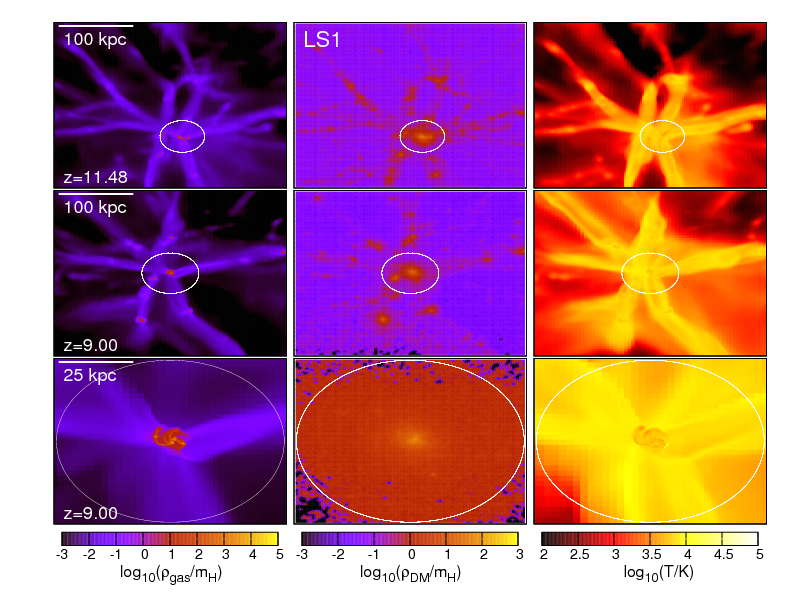}
\includegraphics[width=8.2cm,height=9.0cm]{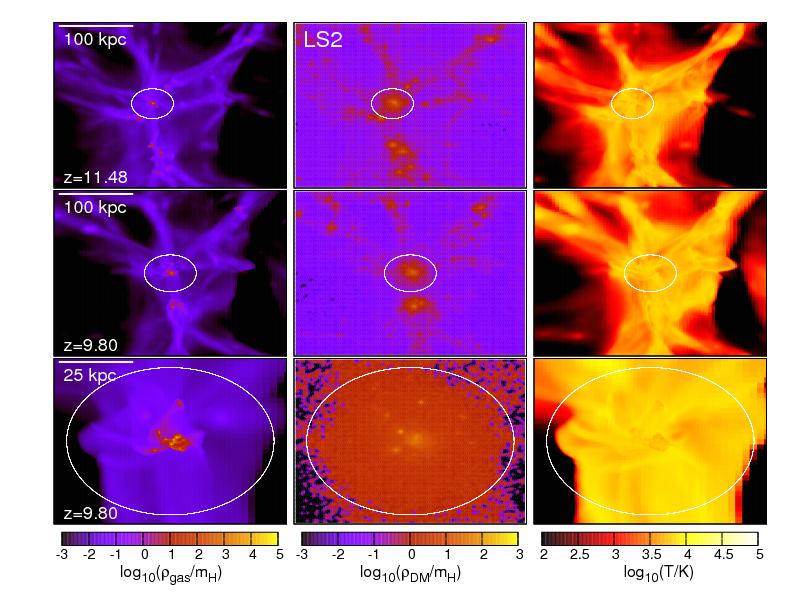}
\includegraphics[width=8.2cm,height=9.0cm]{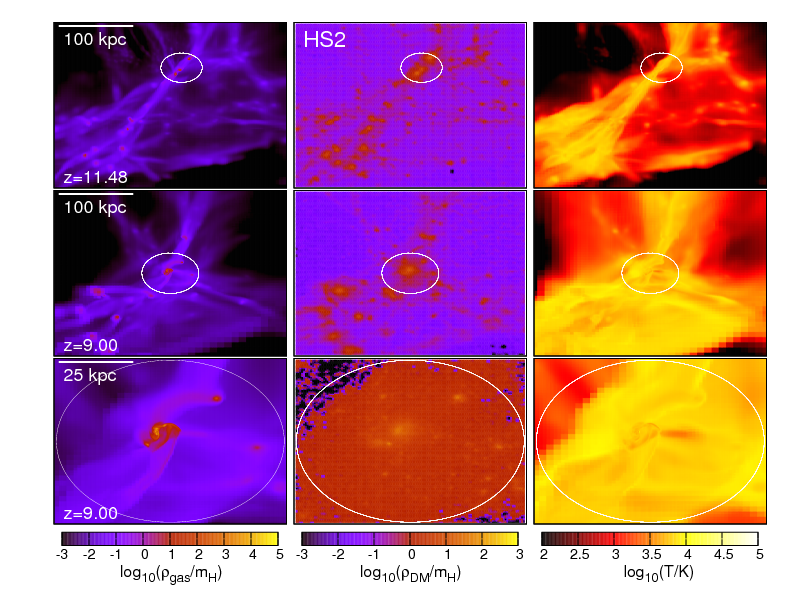}
\includegraphics[width=8.2cm,height=9.0cm]{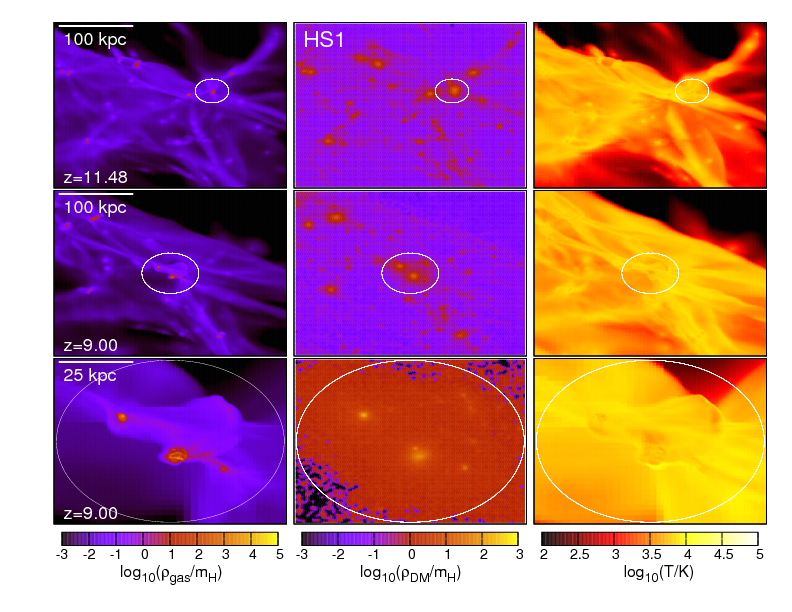}
\caption{Mass weighted gas number density (left), DM number density (center) and gas temperature (right) for the four systems. The white circle shows the virial radius at 
a given redshift. Note the preference for LS systems to be located at center of knots in the cosmic web while LS systems are within filaments (see text for quantitative details).}
\label{fig:haloesmap}
\end{figure*}

The above studies focused on DM haloes with masses ${\rm M_{vir}\ga10^{11}M_{\odot}}$; in order study 
the very initial stages ( ${\rm M_{vir}\sim10^{6}M_{\odot}}$ and larger) of spin acquisition after the turn around we will 
work with haloes in the mass range of ${\rm M_{vir}\approx (1-2) \times10^{9}M_{\odot}}$ at redshift $z\approx10$.

In this paper we want to understand if the spin of primordial galaxies is provided by the dark matter or by random motions of the gas 
within the cosmic web. To this purpose we focused on simulating the properties of atomic cooling  galaxies, which should be 
unaffected by stellar and AGN feedback processes. The possibility that the {\it James Web Space Telescope} ({\it JWST})
can observe the first galaxies, located in haloes $\gsim 10$ times more
massive than those hosting the first PopIII stars, is an exciting
motivation to study the so-called atomic cooling haloes (i. e. DM
haloes with a virial temperature ${\rm T_{vir}\ga10^4K}$; hereafter ACHs
-- see e.g. \citet{Ferrara,BrommYoshida2011} for recent reviews).

\cite{PrietoACH} showed how environment affects the final fate of ACH. Depending on the environment of the 
ACH, it could develop a disk or a very compact object. Motivated by this result, we further 
investigate the role of environment on ACH by studying their spin build up. In particular, we want to 
understand if location along a filament in the cosmic web or inside a knot, leads to different spin values for ACHs. 
It is worth clarifying that ACHs maybe not the dominant galaxy population at high-z as they require a strong UV background 
nearby to suppress H$_2$ formation. While rare, they exists \citep{Fialkov} and thus can be
clean laboratories of primordial galaxies without any effects due to previous stellar formation or feedback mechanisms. 

For this purpose we have performed cosmological numerical simulations that describe the formation 
of ACH in different environments and with different values of the spin parameter ($\lambda$). Our main finding is that 
the final spin value of a halo can be determined by the clustering of the dark matter. In particular, the topology where the 
halo resides determines the efficiency at which the initial (tidal torque originated) spin is destroyed or enhanced; the rate 
and direction of dark matter accretion determines the final fate of angular momentum.

This paper is organized as follows: in section {\S}\ref{Methodology} we
describe the details of the numerical simulations and the halo
sample. In {\S}\ref{results}, we describe our analysis and present the
physical quantities extracted from the simulations. We conclude in 
{\S}\ref{conclusions}.

\section{Methodology and Numerical Simulation Details}
\label{Methodology}

The simulations presented in this work were performed with the cosmological N-body hydrodynamical 
code RAMSES \citep{Teyssier2002}. This code has been designed to study structure formation with 
high spatial resolution using the Adaptive Mesh Refinement (AMR) technique, with a tree-based data 
structure. The code solves the Euler equations with a gravitational term in an expanding universe using 
the second-order Godunov method (Piecewise Linear Method). The cooling module of the code has been 
replaced by the non-equilibrium cooling model for H$+$He primordial gas from the chemo-thermal KROME 
package \citep{jpp,KromePackage}.

Cosmological initial conditions were generated with the mpgrafic code \citep{Prunetetal2008} inside a 
L$=$10 Mpc comoving side box. Cosmological parameters where taken from \citet{Planck2013Results}: $\Omega_m=0.3175$, 
$\Omega_\Lambda=0.6825$, $\Omega_b=0.04899$, $h=0.6711$, $\sigma_8 =0.83$ and $n_s=0.9624$.

\begin{figure}
\centering
\includegraphics[width=1.1\columnwidth]{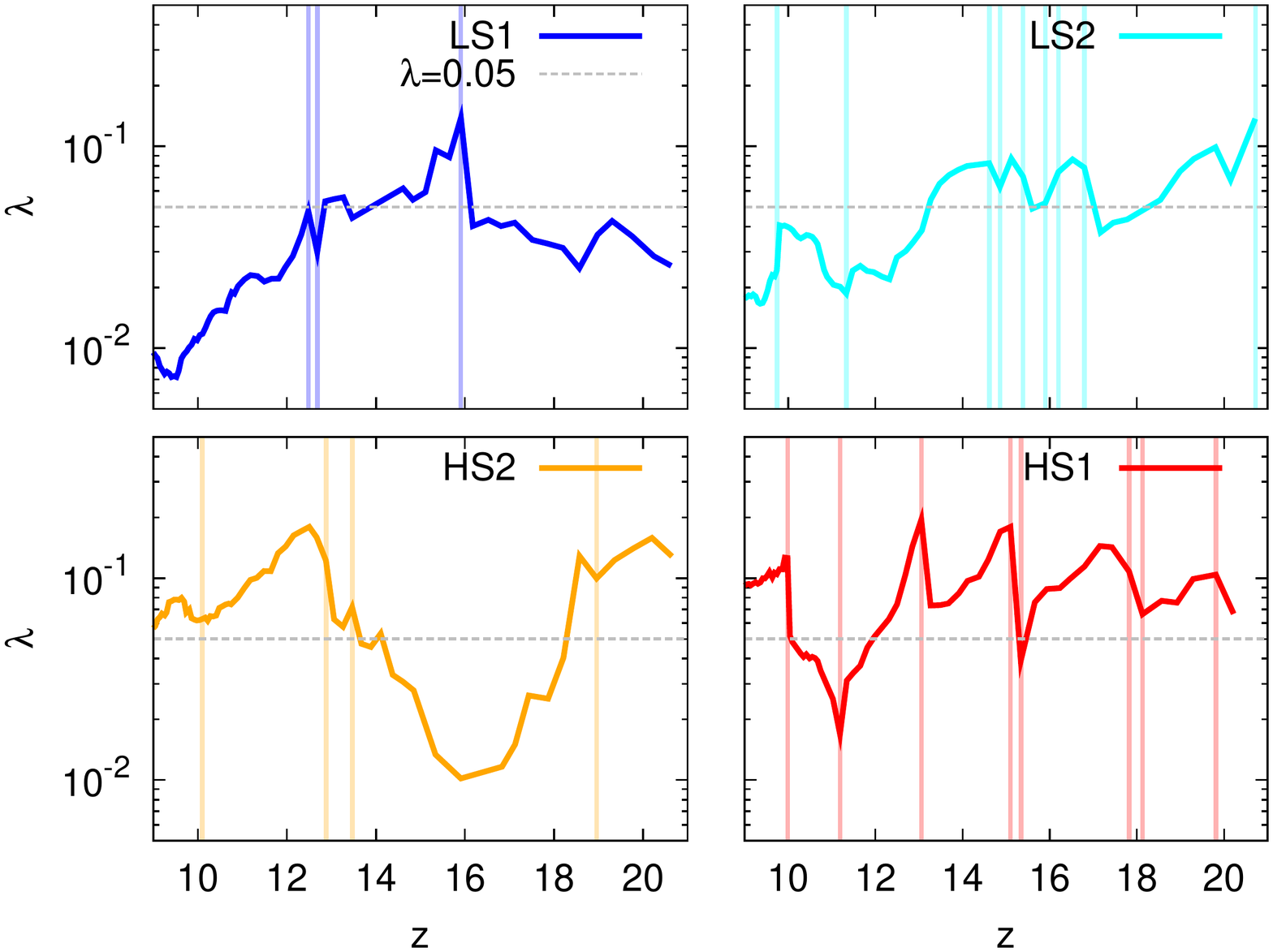}
\caption{Lambda spin parameter as defined in \citet{Bullock2001} for the four systems. We will use the same color convention for the next 
figures in the paper. The dashed gray line is $\lambda=0.05$. Vertical lines show merger events as defined in the text. In general, 
merger events are associated to sharp variations of the spin parameter suggesting 
that these events can play a non negligible role in the DM halo spin acquisition process. The significant decrease and increase in spin for HS2 for $14 < z < 19$ is explained as the transition between isotropic matter accretion (decrease) and the formation of a filament that funnels matter into one preferred direction (increase). See more details in section 3.4 and Fig.~16.}
\label{fig:lambda}
\end{figure}

Using the parameters mentioned above, we ran a number of DM-only simulations with N$_{\rm p}=256^3$ particles 
starting at $z_{\rm ini}=100$. We selected four DM haloes of similar mass M$_{{\rm DM}}\approx 10^9$M$_\odot$ 
and different spin parameter 
\begin{equation}
\lambda \equiv {\rm \frac{J}{\sqrt{2}M_{vir}R_{vir}V_{cir}}}
\end{equation}
(as defined in \citet{Bullock2001}) at redshift $z=10$. Here J, ${\rm M_{vir}}$, ${\rm R_{vir}}$ and ${\rm V_{cir}}$ 
are the DM angular momentum, the halo virial mass, the halo virial radius and the halo circular velocity, respectively.
The halo spin had values in the range $\lambda\approx0.1 - 0.01$. 

After the selection process we re-simulated the haloes including DM and atomic-only H$+$He primordial 
gas physics without star formation. For these simulations we re-centered the box in the DM halo position 
at redshift $z=10$. These simulations were run using 3 DM nested grids filling the whole box. Each of 
these grids had $128^3$, $256^3$ and $512^3$ DM particles (covering the refinement levels from 7 to 9). 
Furthermore, we inserted another DM grid of $512^3$ particles centered at the box center covering 1/8 of 
the total box volume (corresponding to refinement level 10). In this way we were able to reach a DM 
resolution equivalent to a $1024^3$ particles grid inside the box central region, which corresponds to a 
particle mass  m$_{{\rm p}}\approx3\times10^4$M$_\odot$ resolution. 

Following a geometrical refinement strategy, we allowed refinements inside a fixed spherical volume 
centered at the simulation box center. The volume was big enough to properly resolve the whole Lagrangian 
patch finishing inside a radius R$_{{\rm ref}}\approx3$R$_{{\rm vir}}$ at the end of the simulation, 
$z_{{\rm end}}\approx9$.

The refinement criterion was set as follows: a cell is refined if all of the following four conditions are fulfilled i) it 
contains more than 8 DM particles, ii) its baryonic content is 8 times higher than the average in the whole box, iii) the local Jeans length 
is resolved by less than 4 cells \citep{Trueloveetal1997}, and iv) if the relative pressure variation 
between cells is larger than 2. Following these criteria the maximum level of refinement was set 
at $l_{{\rm max}}=16$, corresponding to a co-moving maximum spatial resolution of $\Delta x_{{\rm max}}\approx153$ pc. 
To avoid numerical fragmentation due to the imposed maximum level of refinement we set an artificial temperature 
floor in order to reach the Jeans criterion when a cell at $l_{{\rm max}}$ satisfies one of the refinement conditions. However, due to the relatively high temperature of atomic cooling halos, this floor was never activated in the simulations presented in this paper. The floor generally activates for densities  $> 10^4$ cm$^{-3}$ and temperatures $< 300 K$.

To compute the gas angular momentum we use tracer (gas) particles to follow the Lagrangian patch 
associated to a given gas volume in our four re-simulations. Tracer particles were implemented in
the same way as \citet{Dubois+2012}: for each DM particle there is a tracer particle (of zero mass) at the same position with zero initial 
velocity. It means that at the beginning the tracers have the same distribution of the DM particles. Based on 
the local gas velocity the tracer's position is updated following a simple forward Eulerian scheme. It means 
that at each time step the tracer moves following the local gas behavior. Fig.~\ref{fig:traject} shows the 
trajectories for both DM and tracer particles associated to the patch finishing inside $0.5$R$_{\rm vir}$ at 
the end of the four simulations analyzed in this work (and presented in the next paragraph). From this figure
it is possible to see how the gas trajectories are more concentrated than the DM trajectories. This feature is the
consequence of gas cooling.

\begin{figure*}
\centering
\includegraphics[width=1.0\columnwidth]{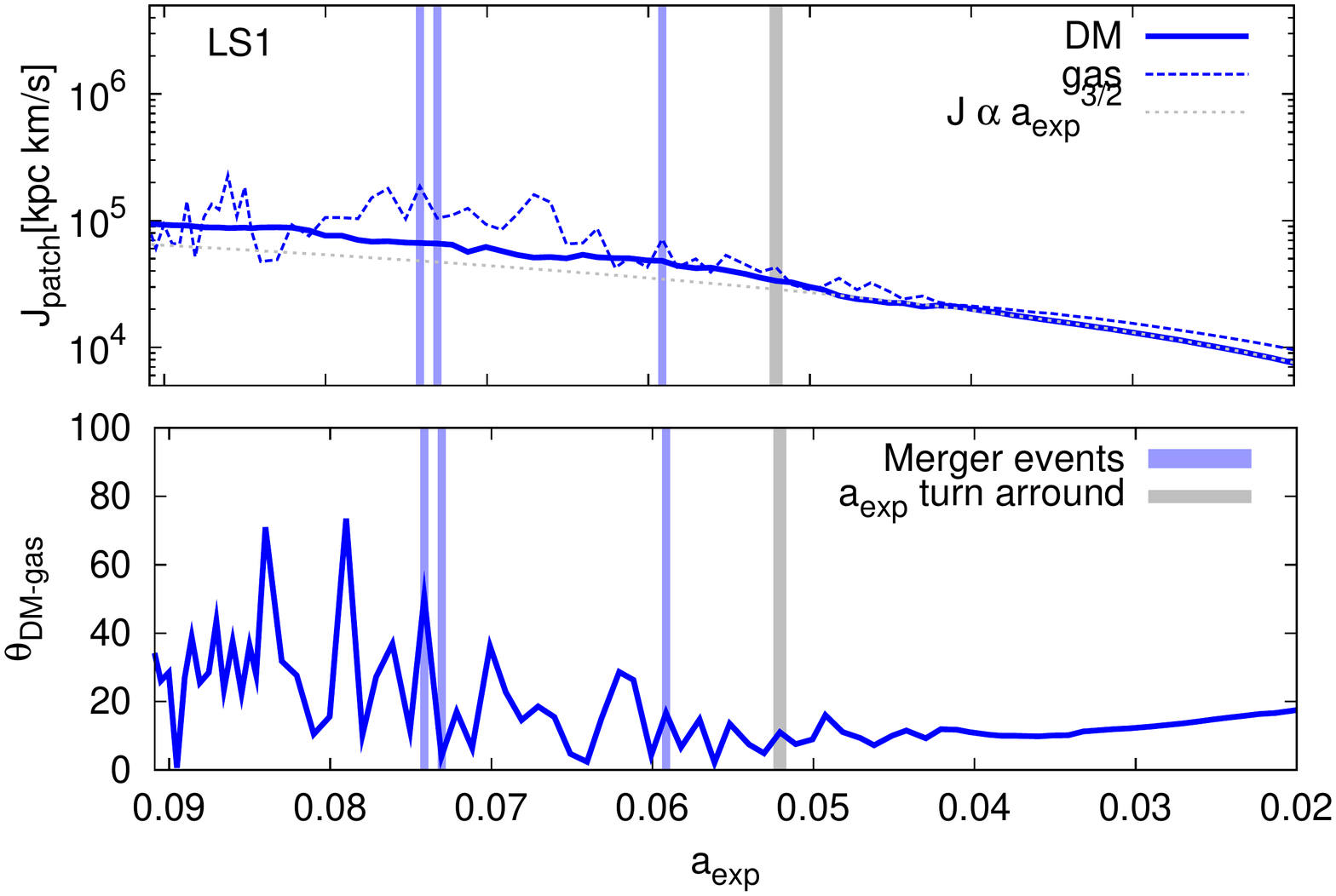}
\includegraphics[width=1.0\columnwidth]{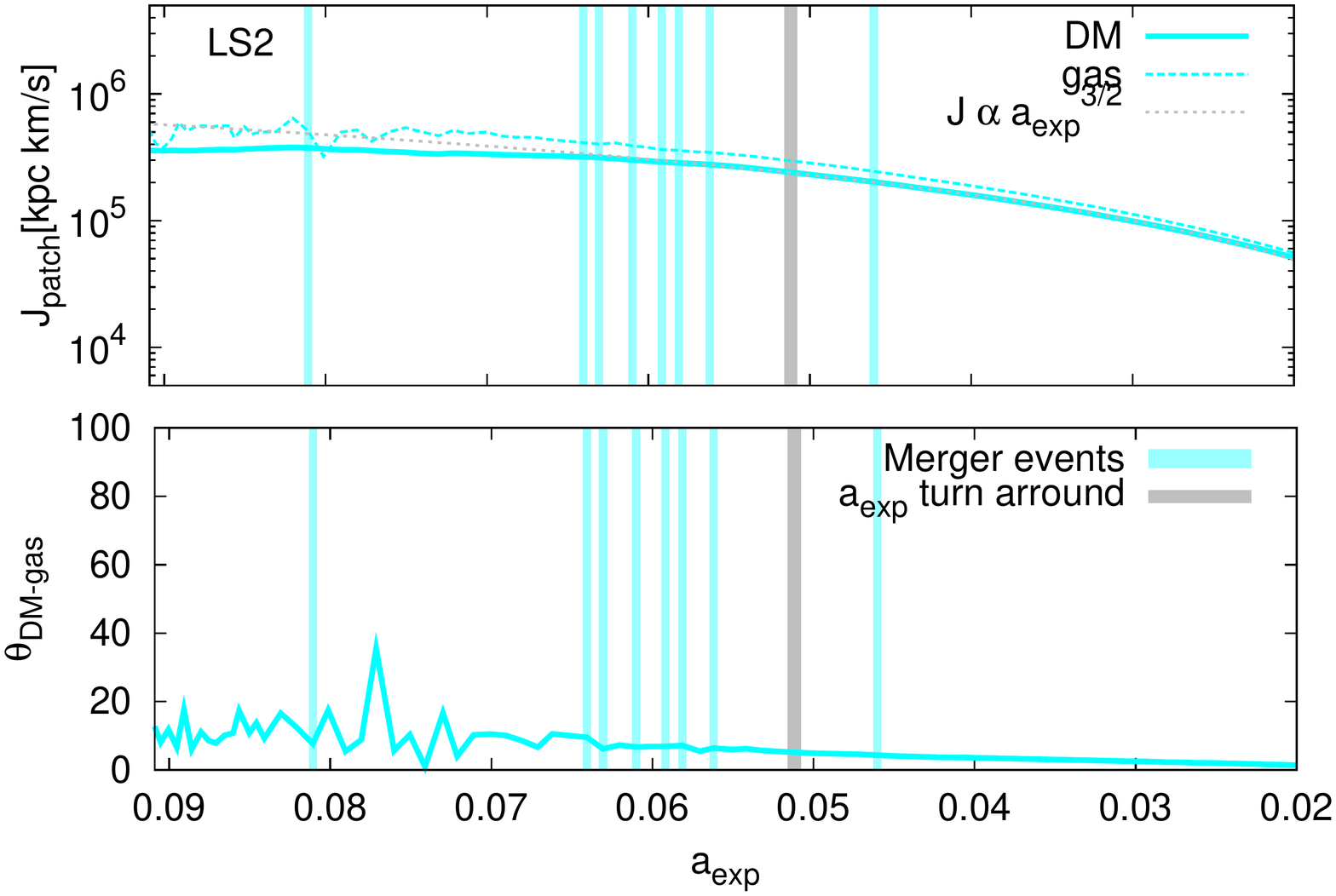}
\includegraphics[width=1.0\columnwidth]{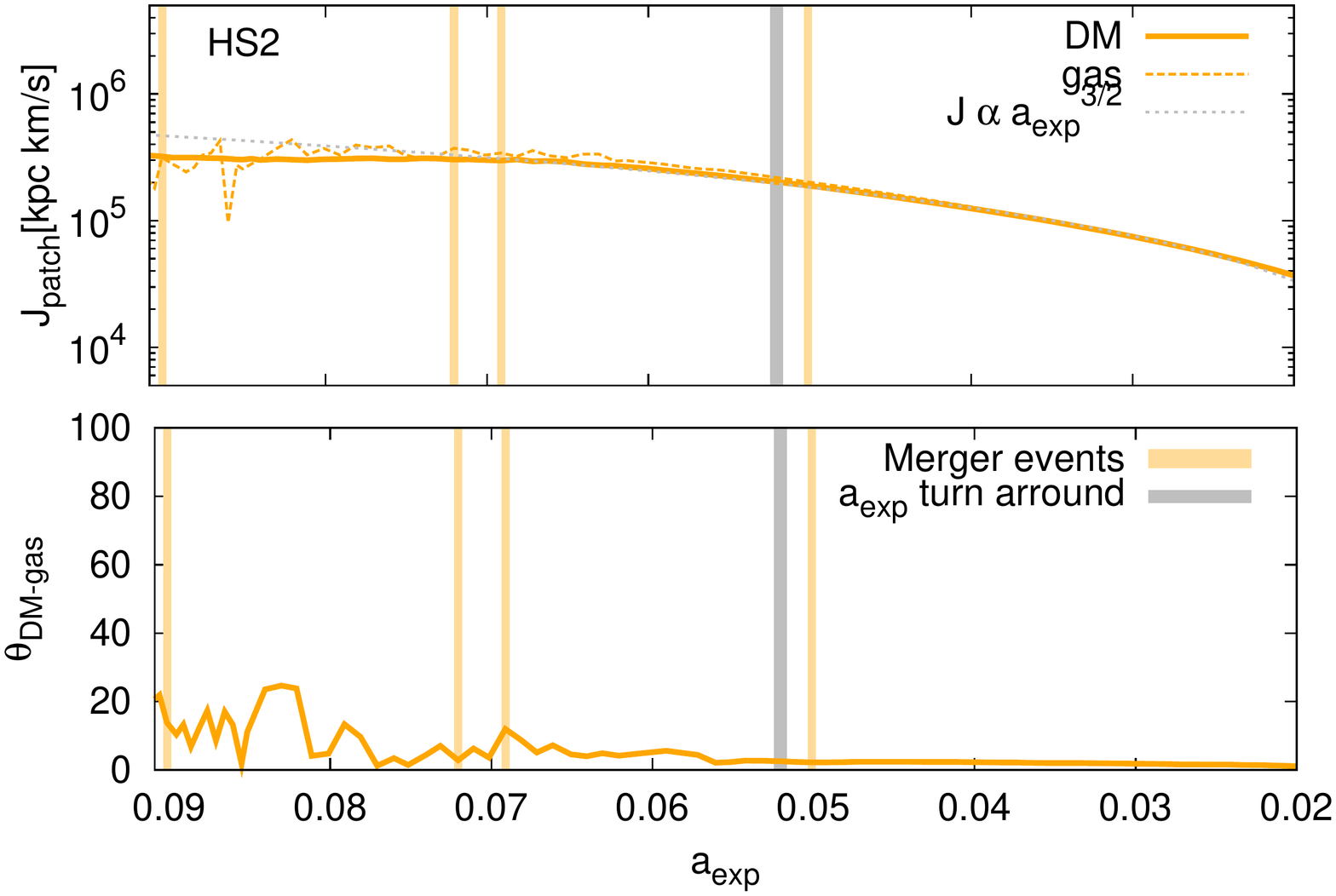}
\includegraphics[width=1.0\columnwidth]{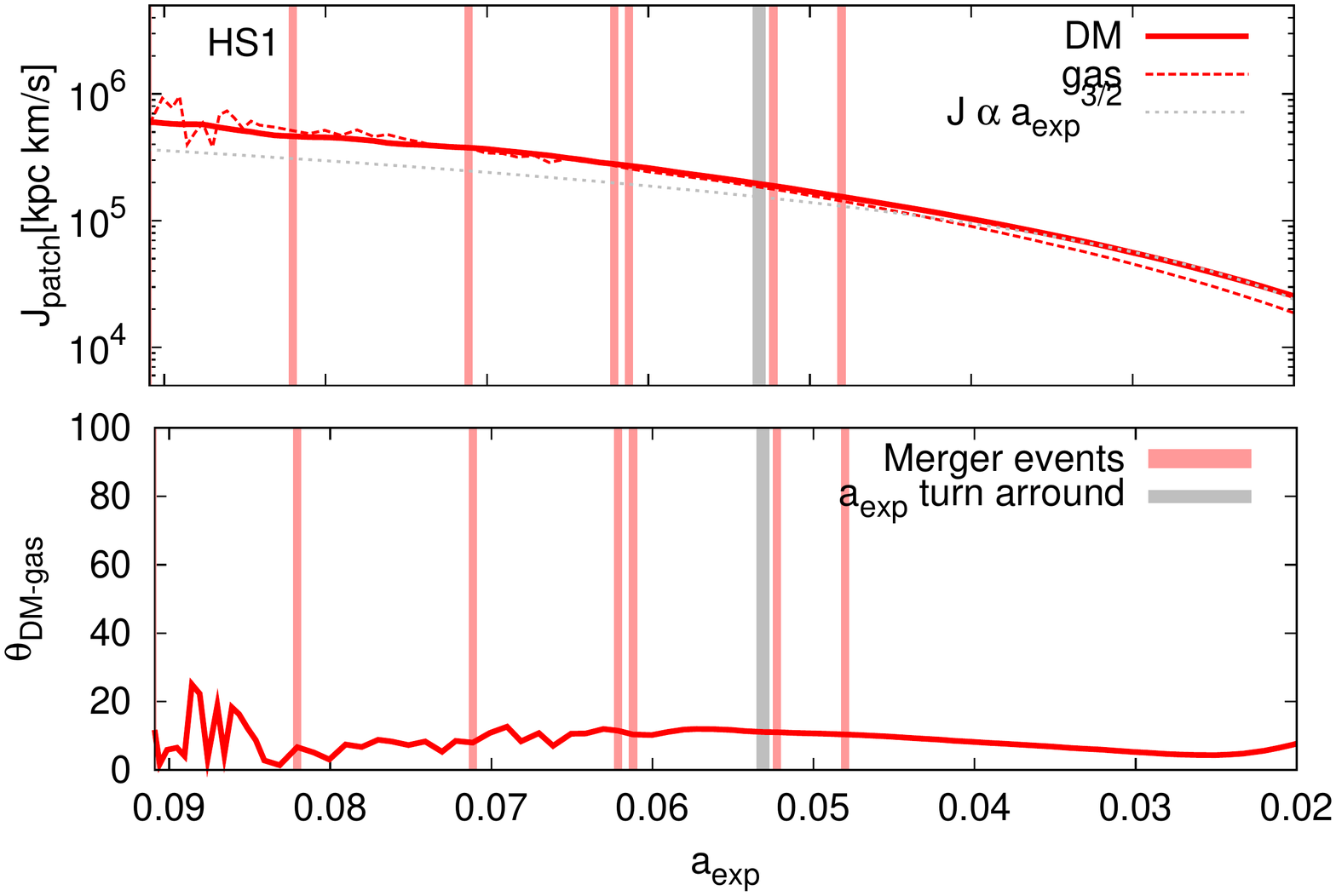}
\caption{Specific angular momentum as a function of $a_{\rm exp}$ for a patch including all the 
particles inside the halo's virial radius at redshift
$z=10$ (top row of each panel). The solid line shows the DM particles evolution and the dashed line shows the gas tracer particles evolution. The dashed
gray line shows the tidal torque theory prediction: $|{\rm J}|\propto a_{\rm exp}^{3/2}$. The blue vertical lines mark the
merger events as defined in the text and the gray vertical line shows the DM patch turn-around.  Misalignment
angle between gas and DM angular momentum vectors as a function of $a_{\rm exp}$ (bottom row of each panel). The figure shows that at the beginning the 
DM patch follows a $|{\rm J}|\propto a_{\rm exp}^{3/2}$ evolution. At this stage the gas component follows roughly the same behavior with
$\theta_{\rm DM-gas}\la20^{\circ}$. In the vicinity of the turn-around point the system deviates from the TTT prediction.
Besides this deviation the gas component starts to show fluctuations in both the SAM and the misalignment angle. Some of these 
fluctuations are directly associated to merger events suggesting that these events are able to misalign the spin of both 
components via pressure gradients. Pressure torques associated to gas accretion are the source for these fluctuations 
between mergers.}
\label{fig:spinpatch}
\end{figure*}

\begin{figure*}
\centering
\includegraphics[width=1.0\columnwidth]{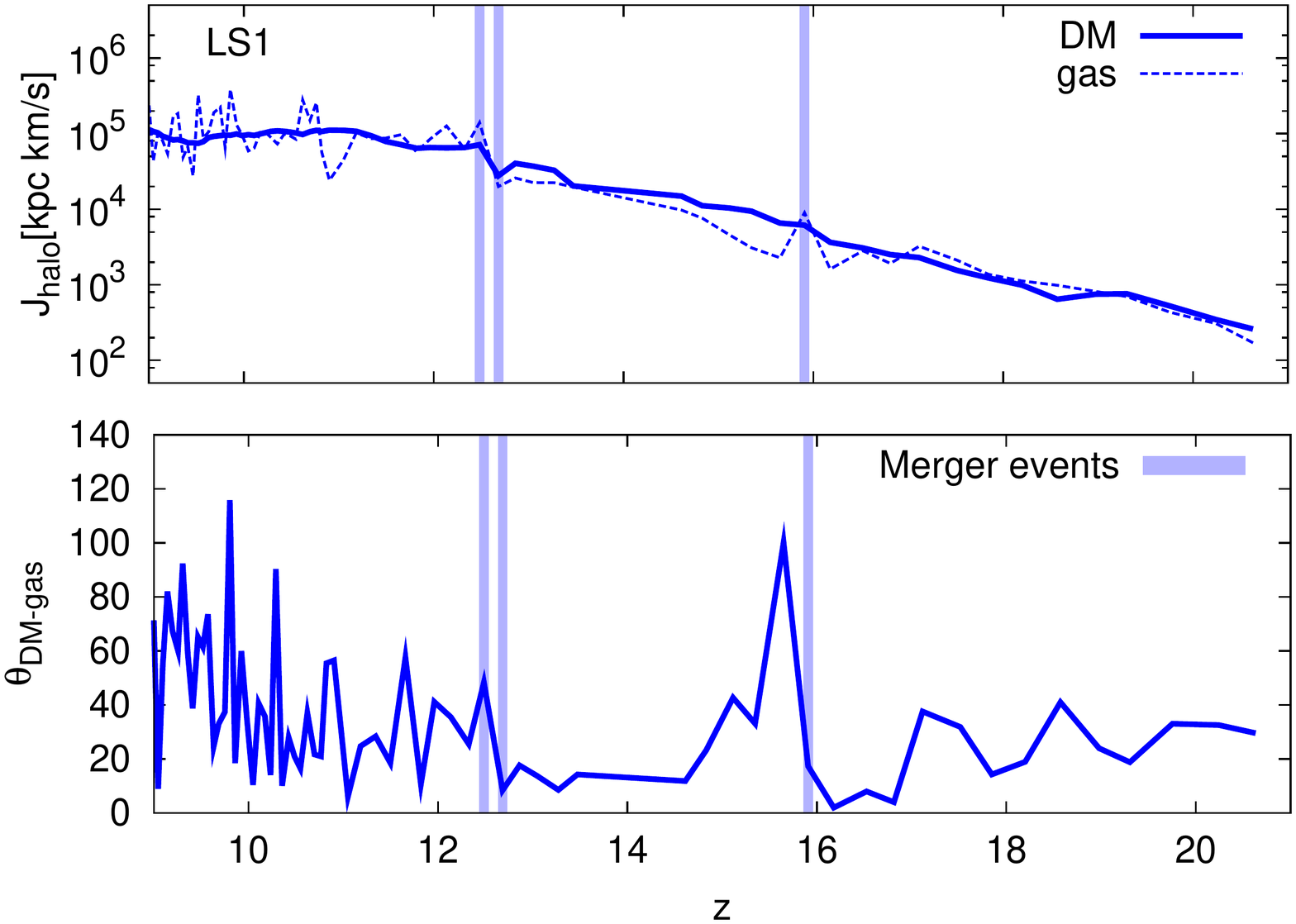}
\includegraphics[width=1.0\columnwidth]{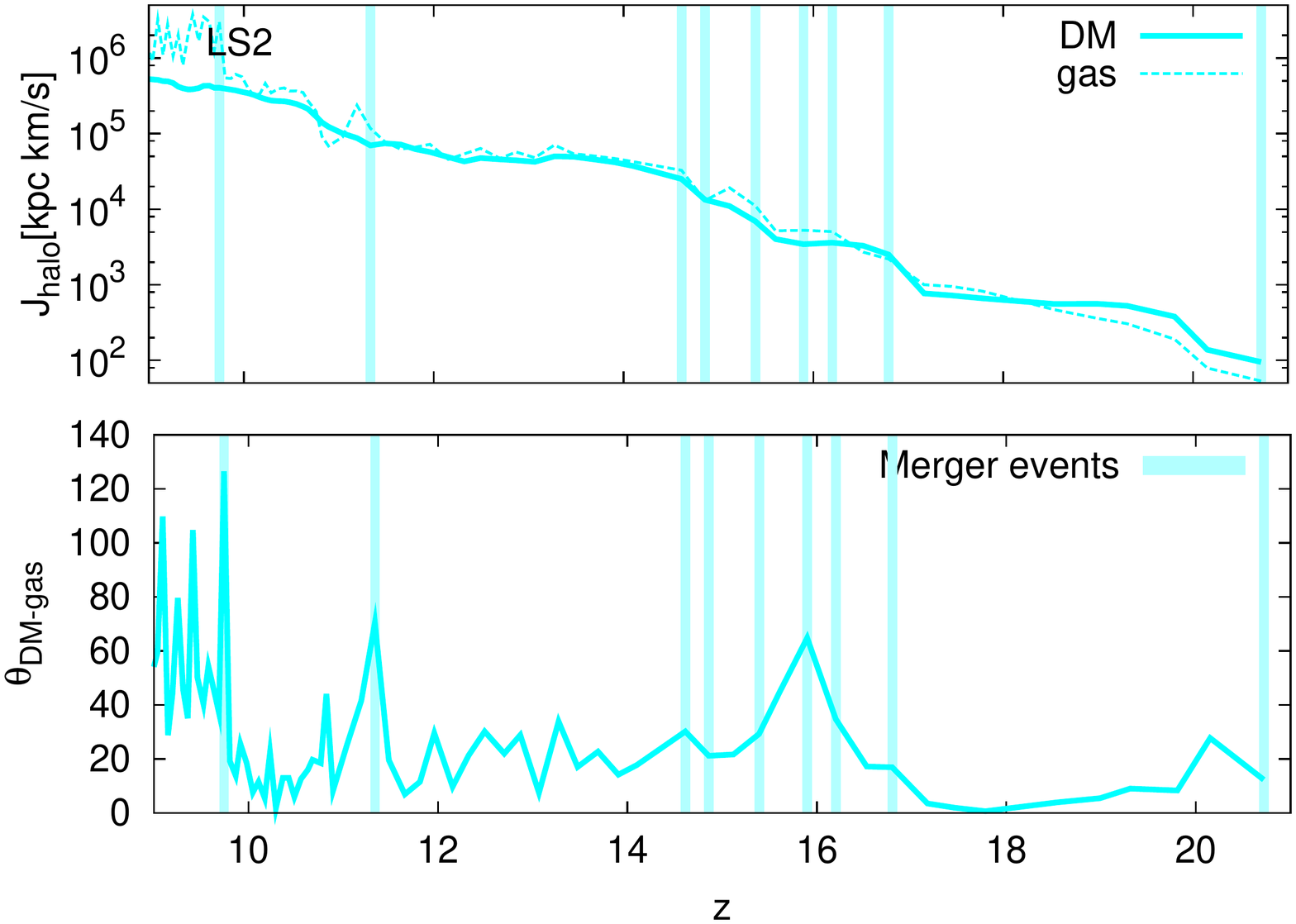}
\includegraphics[width=1.0\columnwidth]{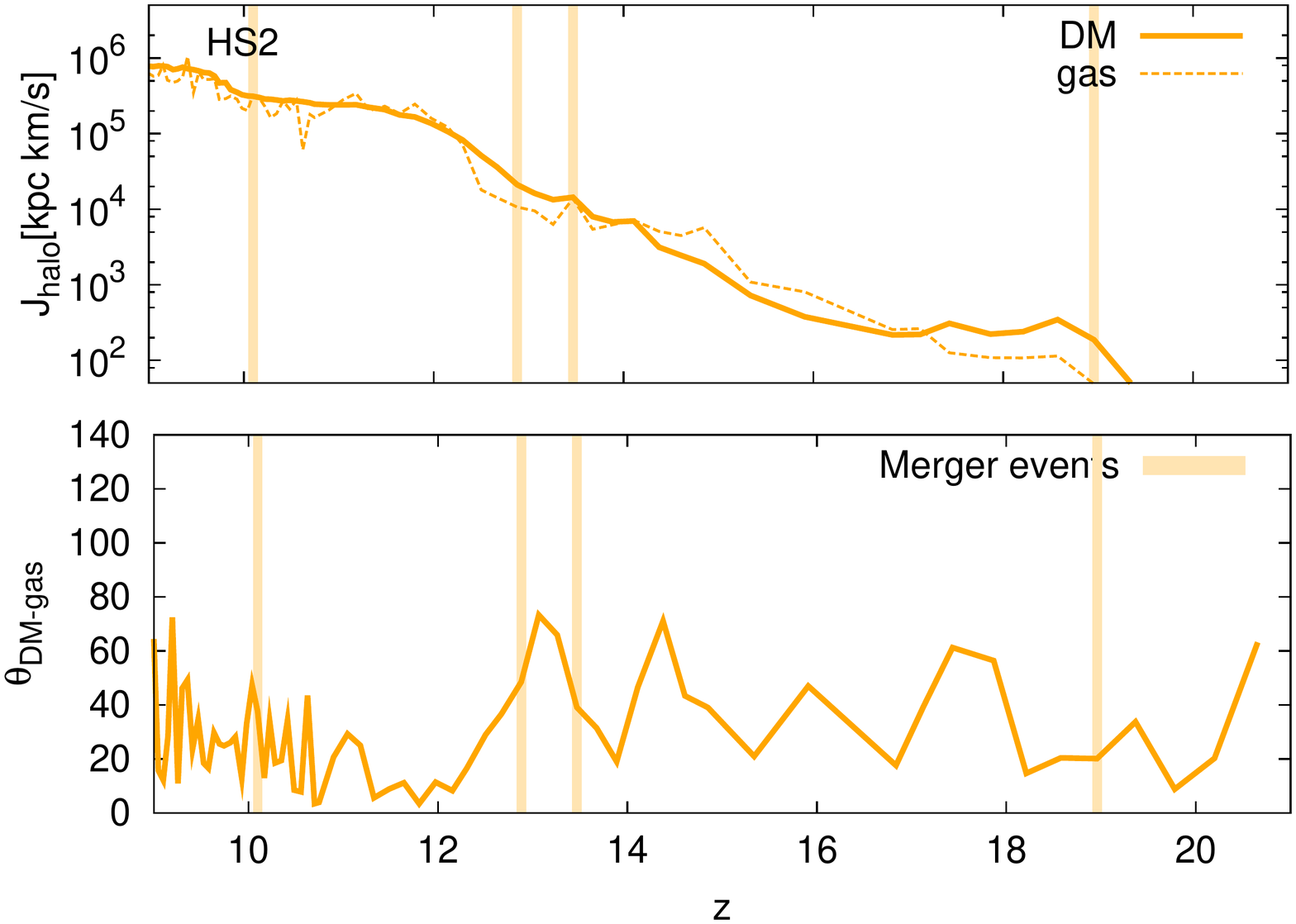}
\includegraphics[width=1.0\columnwidth]{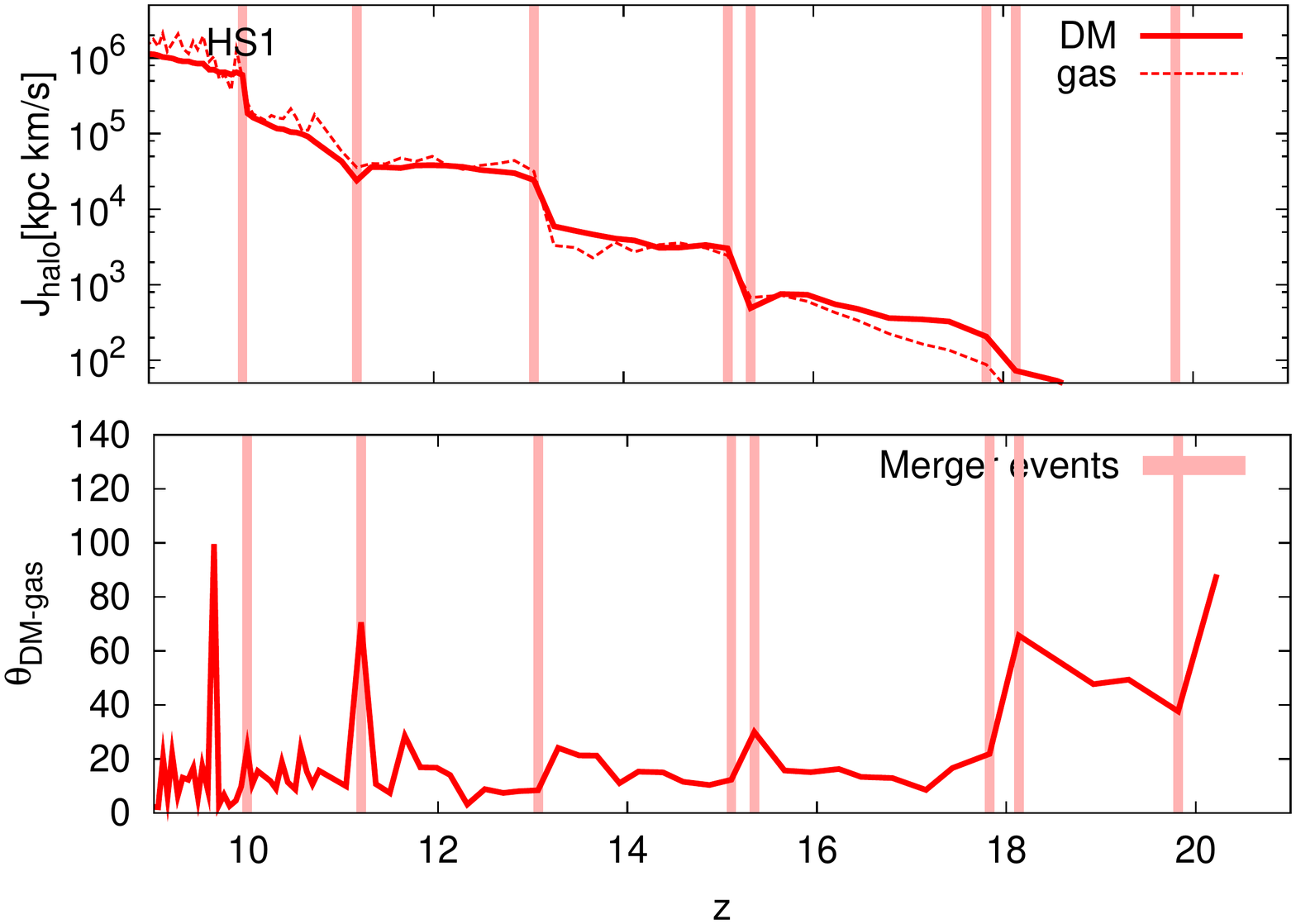}
\caption{Specific angular momentum for the halo as a function of redshift $z$ (top row of each of the four panels). Solid line: DM component
of the system. Dashed line: the gas component of the system. The vertical lines mark merger events as defined in the 
text. Misalignment angle $\theta$ between the spin of both the DM and the gas component of the halo (bottom row of each panel). The figure shows that
the merger events are able to produce strong variations in the misalignment angle. The halo spins
up between mergers suggesting that accreted material plays an important role in the angular momentum acquisition process
(see Fig.~\ref{fig:spinpatchhalo}). Fluctuations of the misalignment angle where there are no mergers suggest that
the accretion process is able to misalign the spin of both components due to pressure torques.}
\label{fig:spinhalo}
\end{figure*}

Based on the spin parameter of the analyzed haloes at redshift $z=10$ we named our four simulations as HS1 
($\lambda\approx0.1$), HS2 ($\lambda\approx0.06$) and LS1 ($\lambda\approx0.01$) and LS2 ($\lambda\approx0.04$) 
where HS refers to High Spin and LS to Low Spin. In order to have an idea of the morphology of our haloes, 
Fig.~\ref{fig:haloesmap} shows the mass weighted (gas and DM) density and gas temperature map for each system. 
The $\lambda$ spin parameter evolution as a function of redshift is shown in Fig.~\ref{fig:lambda}.

\section{Results and Discussion}
\label{results}

\subsection{Initial Spin and Evolution of the Patch.}

We study first the origin and evolution of spin before the virialization stage in order to compare 
it with the prediction from tidal torque theory. Initial angular momentum is acquired via tidal torques acting on the density field, the 
so-called tidal-torque theory (hereafter TTT: \citet{Peebles1969,Doro70,White1984}). TTT states that the initial 
angular momentum of an over-dense region in our Universe comes from tidal torques associated to the non-homogeneous 
gravitational field around it. Quantitatively:

\begin{equation}
{\rm L_i\propto a^2(t)\dot{D}(t)\epsilon_{ijk}T_{kl}I_{lj}},
\label{TTT}
\end{equation}   
where ${\rm L_i}$ is the angular momentum of the region, a(t) is the Universe's expansion scale factor, D(t) 
is the linear perturbation theory growth factor, $\epsilon_{ijk}$ is the Levi-Civita symbol, T$_{kl}$ is the 
tidal field tensor and I$_{lj}$ is the moment of inertia tensor.

For a matter dominated universe, which is a good approximation for our Universe at redshift $z\ga10$, 
the time dependence of the above equation is
\begin{equation}
{\rm a^2(t)\dot{D}(t)\propto a^{3/2}(t)}.
\label{timeTTT}
\end{equation}   

Fig.~\ref{fig:spinpatch} shows the modulus of the specific angular momentum $\vec{{\rm J}}$ (hereafter SAM) as a function of the 
expansion scale factor ${\rm a_{\rm exp}}$ (top panel) and the misalignment angle $\theta$ (bottom panel) between the gas and DM components defined as
\begin{equation}
\cos(\theta)=\frac{\vec{{\rm J}}_{\rm DM}\cdot \vec{{\rm J}}_{\rm trc}}{|\vec{{\rm J}}_{\rm DM}| |\vec{{\rm J}}_{\rm trc}|},
\label{eq:misangle}
\end{equation}
for each of the four simulated haloes. The figure shows the evolution of the patch associated to DM (solid line) and gas tracer 
particles (dashed line) inside R$_{\rm vir}$ at redshift $z=10$. The SAM for both components was computed as
\begin{equation}
\vec{{\rm J}}_{\rm patch}=\sum_{\rm parts. i} (\vec{\rm r}_i-\vec{\rm r}_{\rm DM,CM})\times(\vec{\rm v}_i-\vec{\rm v}_{\rm DM,CM}).
\label{eq:Jpatch}
\end{equation}
In the last expression the sum is applied to each type of particle (DM and tracers) inside the 
patch, $\vec{\rm r}_i$ is the particle position at a given time, $\vec{\rm r}_{\rm DM,CM}$ is the 
DM center of mass position of the patch, $\vec{\rm v}_i$ is the particle velocity and 
$\vec{\rm v}_{\rm DM,CM}$ is the DM center of mass velocity of the patch.

The vertical lines mark the merger events suffered by the main halo inside the patch. We only show merger 
events for which the mass ratio M$_2/$M$_1\geq 0.1$, where M$_2$ is the mass of the secondary 
progenitor of the halo and M$_1$ is the mass of the main halo progenitor. Even if these merger events do not have a 
significant effect on the patch evolution they will be very relevant for the main halo evolution as will be 
shown in the next sub-section.

The grey dashed line in the figure shows $|\vec{{\rm J}}_{\rm patch}|\propto a^{3/2}(t)$. It is clear that both DM and 
gas tracer particles follow  TTT predictions up to redshift $z\approx20$. This value is in the 
vicinity of the turn-around redshift of the patch which is marked by a vertical grey line. After this 
redshift the patch's SAM starts to increase at a slower rate than before turn-around till it reaches 
an almost constant maximum value in the redshift range $z\approx11-10$. After turn-around the gas SAM starts to fluctuate in short time scales. Such fluctuations are not present in the 
DM SAM evolution. Because both components feel the same gravitational potential the most plausible explanation 
for this different behavior is the combination of cooling and collisional nature of the gas which after shell crossing is able to create 
torques associated to the gas gradient pressure, ${\rm \vec{r}\times\nabla P}$.

The value of the misaligned angles showed in Fig.~\ref{fig:spinpatch} are $\theta<20^\circ$ 
before turn-around. Such low values show that both gas and DM components tend to be aligned in the first stages of the patch evolution, 
i.e. both DM and gas feel the same gravitational potential around them and start 
to rotate in a similar way. In the vicinity of turn-around the misalignment angle starts to fluctuate 
showing variations at time scales much shorter than those associated to the DM SAM. The main source of 
such fluctuations is the non-linear effects associated to the gas pressure gradient on the gas patch SAM
as we will show in the next section. 

\subsection{Spin Evolution of the Main Halo.}

After having analyzed the evolution of the SAM associated to the Lagrangian patch of the DM halo at $z=10$, we study 
the behavior of the SAM associated to the material inside the virial radius of the main 
halo as a function of redshift. This quantity was computed using eq. \ref{eq:Jpatch} with the sum extended to all 
particles inside the virial radius of the halo at a given redshift. 

Fig.~\ref{fig:spinhalo} shows the same quantities as in Fig.~\ref{fig:spinpatch} but for the main halo as a 
function of redshift. From the four panels of this figure we can see that the SAM modulus of both DM  and gas 
follow the same trend but the gas component exhibits greater fluctuations than the DM evolution. 

We computed the misalignment angle between 
the gas and  DM SAM inside the virial radius using eq. \ref{eq:misangle}. In contrast 
with the result of Fig.~\ref{fig:spinpatch}, for the single halo SAM  it is possible 
to see a correlation between some merger events (marked by vertical lines in Fig.~\ref{fig:spinhalo}) 
and the misalignment angle of the different components. Such correlations are present in the four systems 
and are the manifestation of pressure gradients working on the gas due to the violent interaction 
between the dense central regions of the halo progenitor. Besides the fluctuations associated to 
occasional merger events, there are fluctuations between the vertical lines also. Such fluctuations are 
associated to pressure torques triggered by soft accretion (including mergers with a mass ratio M$_2/$M$_1<0.1$)
into the halo central region.

\begin{figure}
\centering
\includegraphics[width=1.1\columnwidth,height=10.0cm]{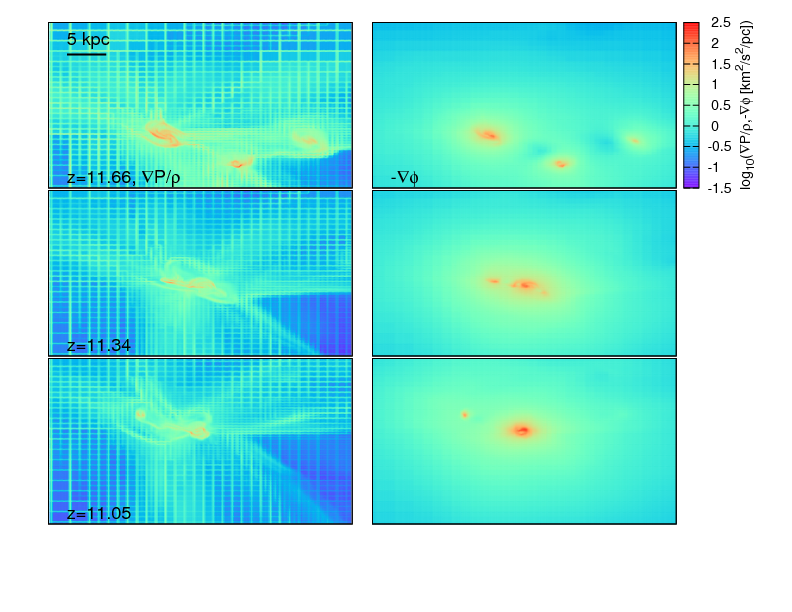}
\caption{Acceleration associated to pressure gradient (left) and gravity gradient (right) maps for a merger event for halo LS1 before, during and after the merger. 
Pressure gradient which only the gas feels are comparable in magnitude to gravity gradients, which both DM and gas feel. These pressure gradient (and its associated torque), 
due both to cooling of the gas and density enhancements due to turbulent motion, provide an extra source of spin generation for the gas.}
\label{fig:pressuregrad}
\end{figure}

Fluctuations in the gas SAM are originated by gas pressure torques which depend on the temperature and density, which are strongly modified by the cooling. Since there is a 
temperature floor (${\rm T \sim 10^4 K}$) for the gas, the gas gets denser and the pressure increases, in other words, isothermal collapse will always put the gas onto a higher adiabat, 
compared to a fixed adiabat with no cooling. It is therefore natural that cooling will change the dominant torque, increasing it given that pressure gradients will increase. This 
torque (which does not affect the DM evolution) can decouple the gas evolution from the DM evolution creating the short time scale variations in the gas SAM. To quantify the role 
of pressure torques at causing fluctuations and spin angle misalignment in the gas component, we compute the acceleration associated to the pressure gradient for a typical merger 
event and compared it to the gravitational acceleration associated to the gravitational potential gradient, in particular we computed
\begin{equation}
{\rm a_{gas}= \frac{| \nabla P |}{\rho} \,\,\,\,\,  {\rm versus}  \,\,\,\,\, a_{grav} = |-\nabla \Phi |}.
\end{equation}
Fig.~\ref{fig:pressuregrad} shows this before, during and after a merger for the LS1 halo. Pressure gradients (and its associated torque)  are comparable to gravitational ones in the high 
density regions. Within our analysis it is not possible to determine if the pressure torques are dominated by cooling of the gas or by the collisional nature of the gas that after shell 
crossing provides pressure induced torques. We will investigate this in an upcoming publication.

\begin{figure}
\centering
\includegraphics[width=1.0\columnwidth]{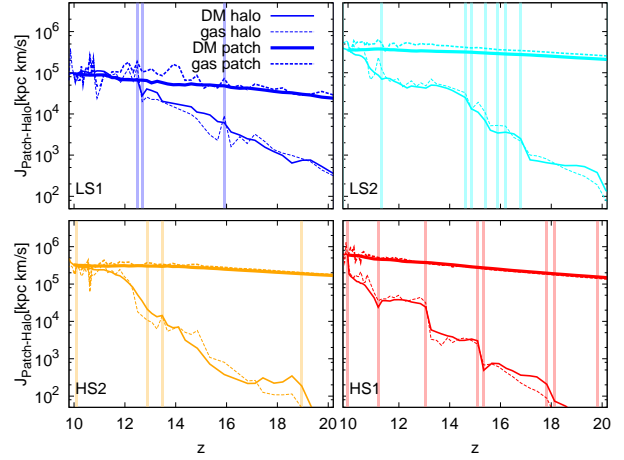}
\caption{Evolution of both the main halo specific angular momentum (thin line) and the specific angular momentum of the patch
associated to the main halo at redshift $z=10$ (thick line) for each of the four systems. Solid line: DM component and in dashed line is the gas component
of the systems. From this it is possible to see that the halo's final ($z=10$) specific angular
momentum comes from two sources: i) mergers, indicated by vertical lines and ii) accretion between merger events. The lowest $\lambda$ halo (LS1) shows a main halo spin evolution with a roughly constant slope 
barely disturbed by 3 merger events. This behavior suggests that in this system most of the angular momentum comes from
soft accreted material. On the other hand, the highest $\lambda$ halo (HS1) shows a staircase-like (far from constant slope)
evolution where each of the steps is associated to merger events. The other two cases, LS2 and HS2, show a mixed behavior
with periods of constant slope evolution and jumps associated to a stair like (mergers) evolution.}
\label{fig:spinpatchhalo}
\end{figure}

It is interesting to have a direct comparison between the two analysis presented above, i.e. a patch v/s main halo evolution.
Fig.~\ref{fig:spinpatchhalo} shows the SAM evolution for the four systems (for gas and DM). The thick lines correspond to 
the patch evolution and the thin lines to the main halo evolution. Our two extreme spin cases, namely LS1 (top-left) and HS1 (bottom-right), have very different behaviors. 
Whereas the LS1 halo shows three merger events barely disturbing the SAM value and a soft mass accretion producing an almost constant
slope in the main halo SAM evolution, the HS1 halo presents merger events which drastically change the SAM value. These 
changes in the SAM value (and in the $\lambda$ evolution, see figure Fig.~\ref{fig:lambda}) are captured as steps in the figure, a feature that does not appear in the LS1 halo. 
Such  step-like structure shows that in this case  mergers contribute significantly to the SAM of the final halo. The other
two intermediate cases (LS2 and HS2) show a mixed behavior with periods of constant SAM slope and step-like evolution due to mergers.

\subsection{Accretion and Mergers.}

After showing how the main halo builds up its SAM from merger events and mass accretion it is worth to study these
processes in more detail. In order to do so we computed the mass accretion history of the haloes, the SAM flux into each halo
and the SAM acquisition from accreted material and the merged secondary halo for a single merger event. 
Together, all these quantities allow us to infer how the haloes at high redshift acquire their spin.

\begin{figure}
\centering
\includegraphics[width=1.0\columnwidth]{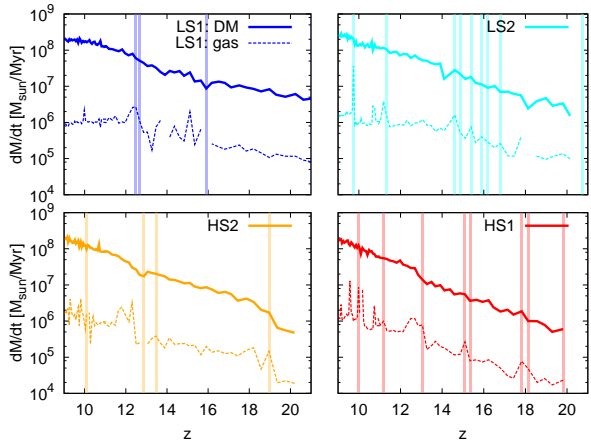}
\caption{Mass accretion evolution. DM evolution (solid line) and gas evolution (dashed line).}
\label{fig:MassAcc}
\end{figure}

\begin{figure}
\centering
\includegraphics[width=1.0\columnwidth]{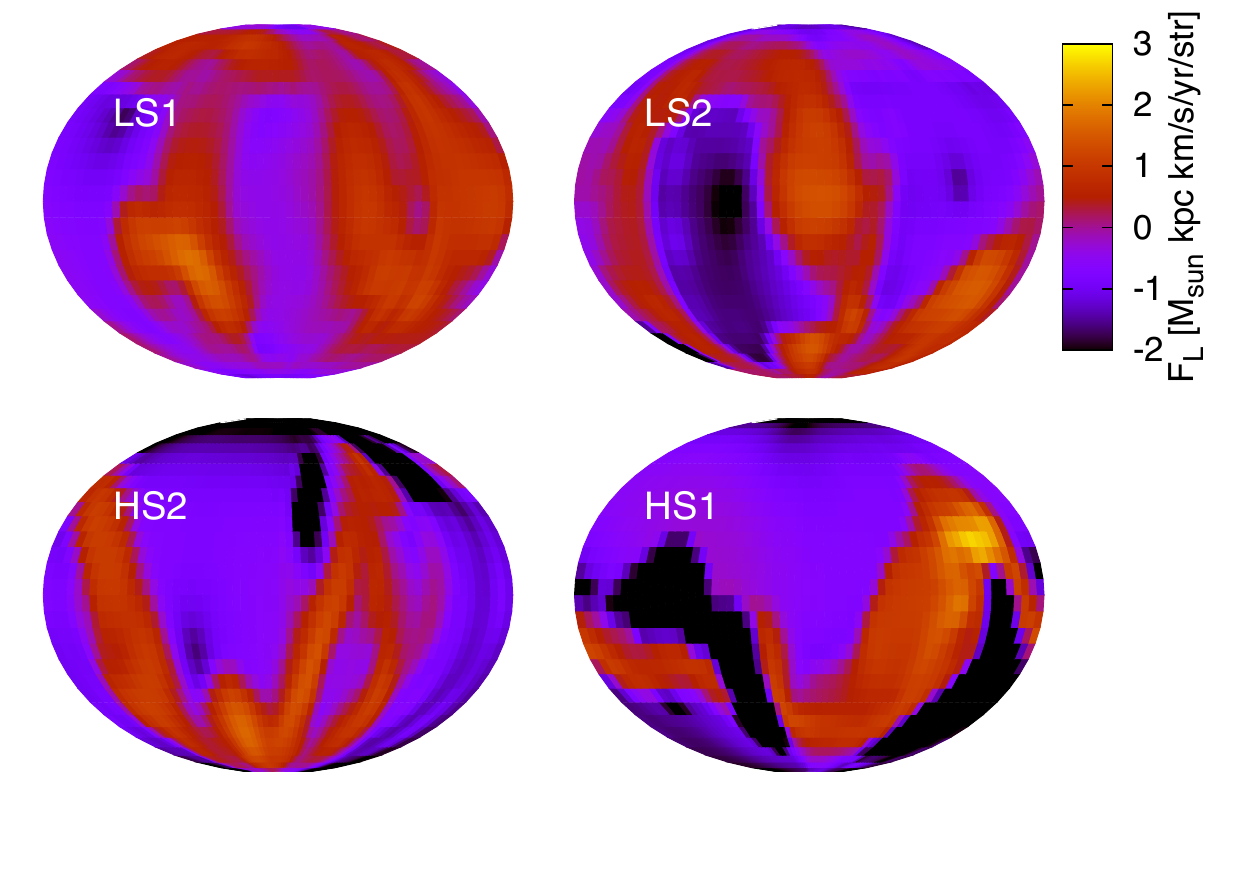}
\includegraphics[width=1.0\columnwidth]{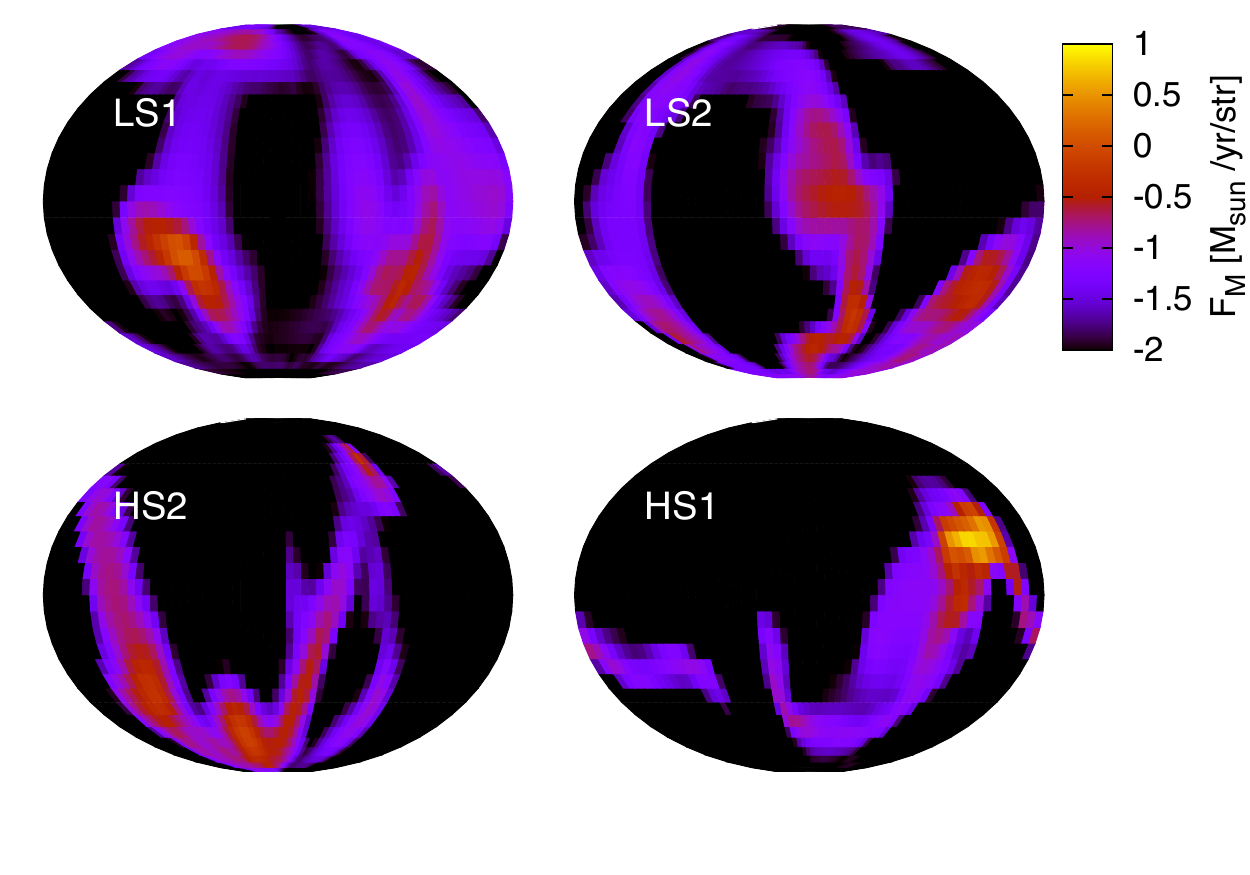}
\caption{Mollweide projections. Top: mass weighted angular momentum flux for each of the four systems. For the low spin systems the angular momentum flux is more isotropic than for the high-spin system, which is more concentrated along a preferred axis. The difference between LS1 and HS1 is especially striking. Bottom: mass weighted mass accretion for each of the four systems. High angular momentum flux regions match to the high mass flux regions. Such a relation suggests that most of the system's angular momentum inside the sphere of 0.5R$_{\rm vir}$ is coming from the accreted mass.}
\label{fig:flux}
\end{figure}

We start with the accretion features in our four systems. Fig.~\ref{fig:MassAcc} shows the gas and DM mass accretion 
history for each of our four haloes. The mass accretion at a given redshift was computed as the ratio
\begin{equation}
\frac{d{\rm M}(z({\rm t}))}{d{\rm t}}=\frac{{\rm M}_{\rm vir}({\rm t}+\Delta {\rm t})-{\rm M}_{\rm vir}({\rm t})}{\Delta {\rm t}}
\end{equation}

From this figure it is not easy to find dramatic differences between the four systems. The most clear difference is associated
to the merger history of the haloes. The haloes with more mergers tend to have more peaks in the gas accretion evolution.
This feature is directly associated to the large amount of mass accreted in a short time step.
Another difference shows up when we compare the DM mass accretion of the LS1 halo with the other three systems. 
In the lowest spin halo (LS1) the DM accretion evolution shows a slower slope compared with the others\footnote{Due to 
the definition based on 
the virial mass, there are some points with no data in the figure. This is because in some steps the halo defined by the halo 
finder keeps its mass constant or it is slightly reduced.}. 

Fig.~\ref{fig:flux} shows Mollweide projections of the distribution of mass weighted in-falling gas angular momentum flux (in units of M$_{\odot}$
kpc km/s/yr/str) at $z=10$ for the four different haloes at the top panel and the mass accretion flux (in units of M$_{\odot}$
/yr/str) at the bottom panel. The angular momentum flux was computed as
\begin{equation}
{\rm F_{{\rm L}}}=\sum_{\rm shell(\Delta r)} \rho {\rm r^2|\vec{r}\times(\vec{v}-\vec{v}_{CM.gas})|\hat{r}\cdot(\vec{v}-\vec{v}_{CM.gas})},
\end{equation}
where the sum was performed including all cells at position ${\rm \vec{r}}$ inside a spherical shell of radius 
${\rm |\vec{r}|\equiv r=0.5 R_{vir}}$ and thickness $\Delta {\rm r}=4\Delta {\rm x}$, with $\Delta {\rm x}$ the cell side at 
the maximum level of refinement; ${\rm \hat{r}}$ is the unitary vector in the ${\rm \vec{r}}$ direction. In the same way the 
mass flux was computed as
\begin{equation}
{\rm F_{{\rm mass}}}=\sum_{\rm shell(\Delta r)} \rho {\rm r^2\hat{r}\cdot(\vec{v}-\vec{v}_{CM.gas})}.
\end{equation}

From the figure (bottom panel) 
it is possible to see that 
the two fast rotating haloes HS2 and HS1 tend to acquire angular momentum flux from a preferred hemisphere, from south and 
north, respectively. Another interesting remark is that both LS1and LS2 are  acquiring angular momentum  from nearly all directions 
without a preferred angle. From the bottom panel it is possible to see that the high angular momentum flux regions are good matched to the high mass flux regions. Such a correlation suggests that most of the angular momentum accreted on the halo's central region
is coming with the accreted mass (as has been seen in more massive systems by \citet{Powelletal2011} and \citet{Dubois+2012}). 

\begin{figure*}
\centering
\includegraphics[width=11.0cm,height=22.0cm]{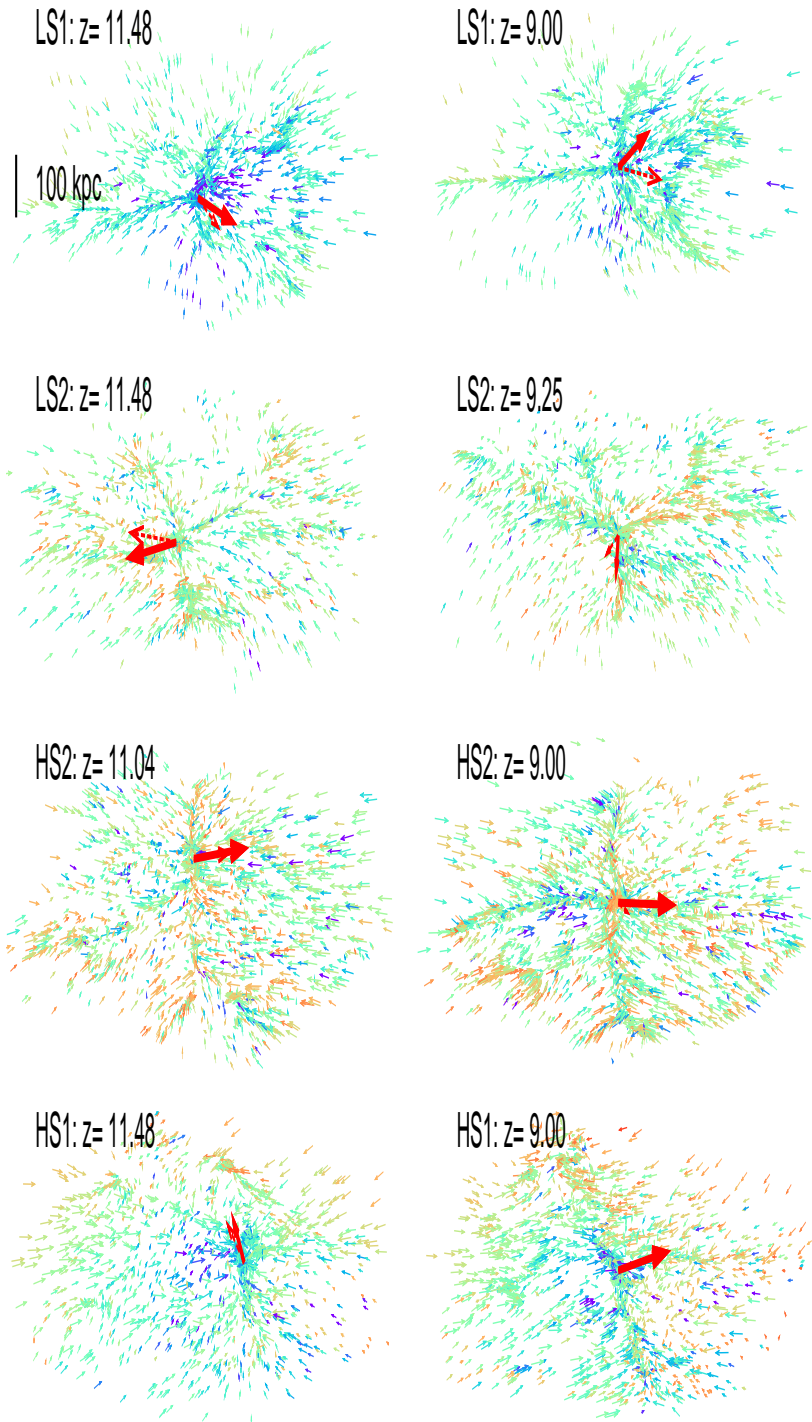}
\caption{Tracer particle's velocity vector field for the low spin (top) and the high spin systems (bottom) at 
two different moments. The different colors show the specific angular momentum of each particle at $z\approx70$. 
Red indicates high values and magenta low values. The red arrows show the specific angular momentum of the main central 
halo for gas (in dashed line) and DM (in solid line).}
\label{fig:vectorfield}
\end{figure*}

\begin{figure*}
\centering
\includegraphics[width=1.0\columnwidth]{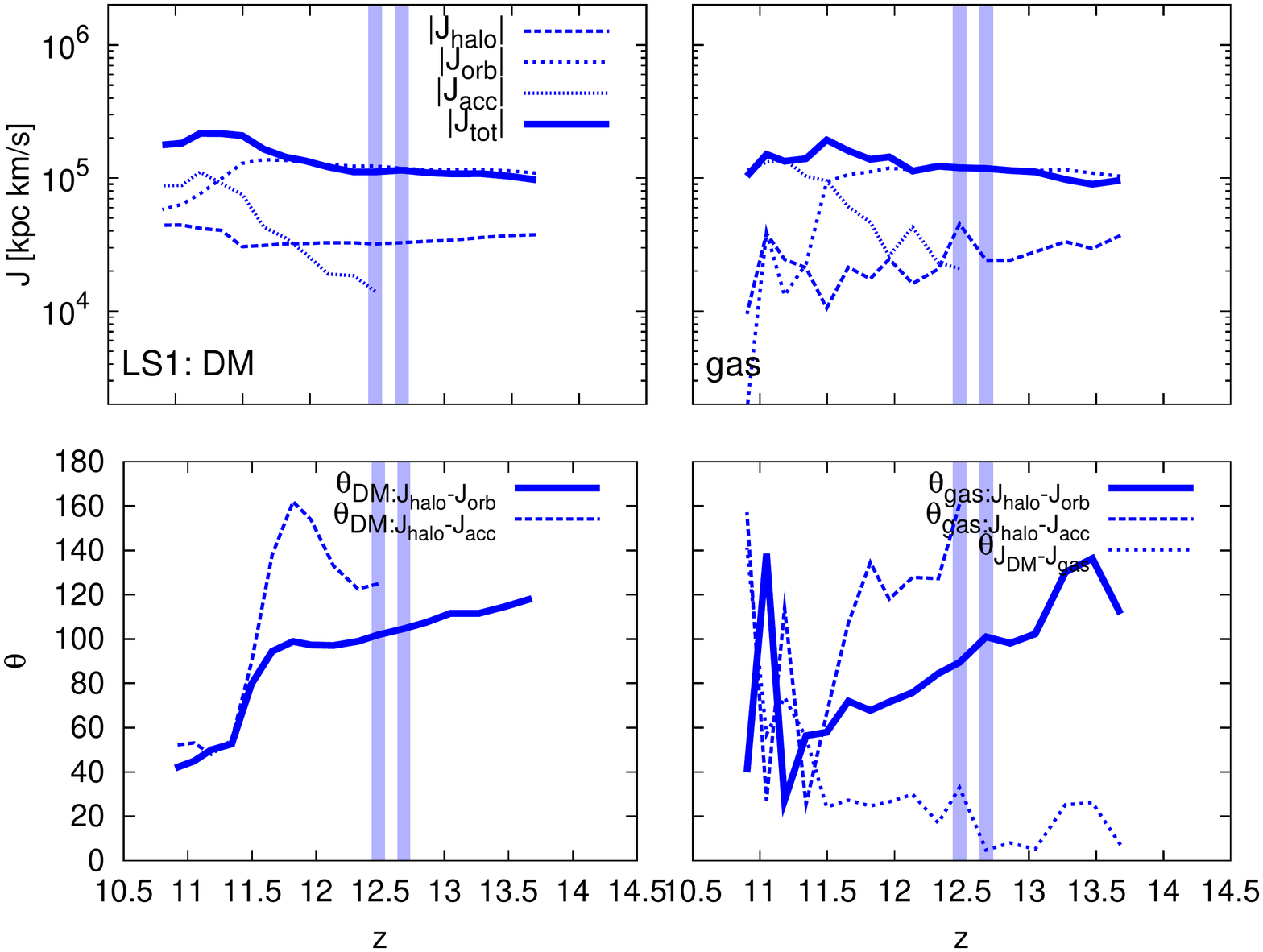}
\includegraphics[width=1.0\columnwidth]{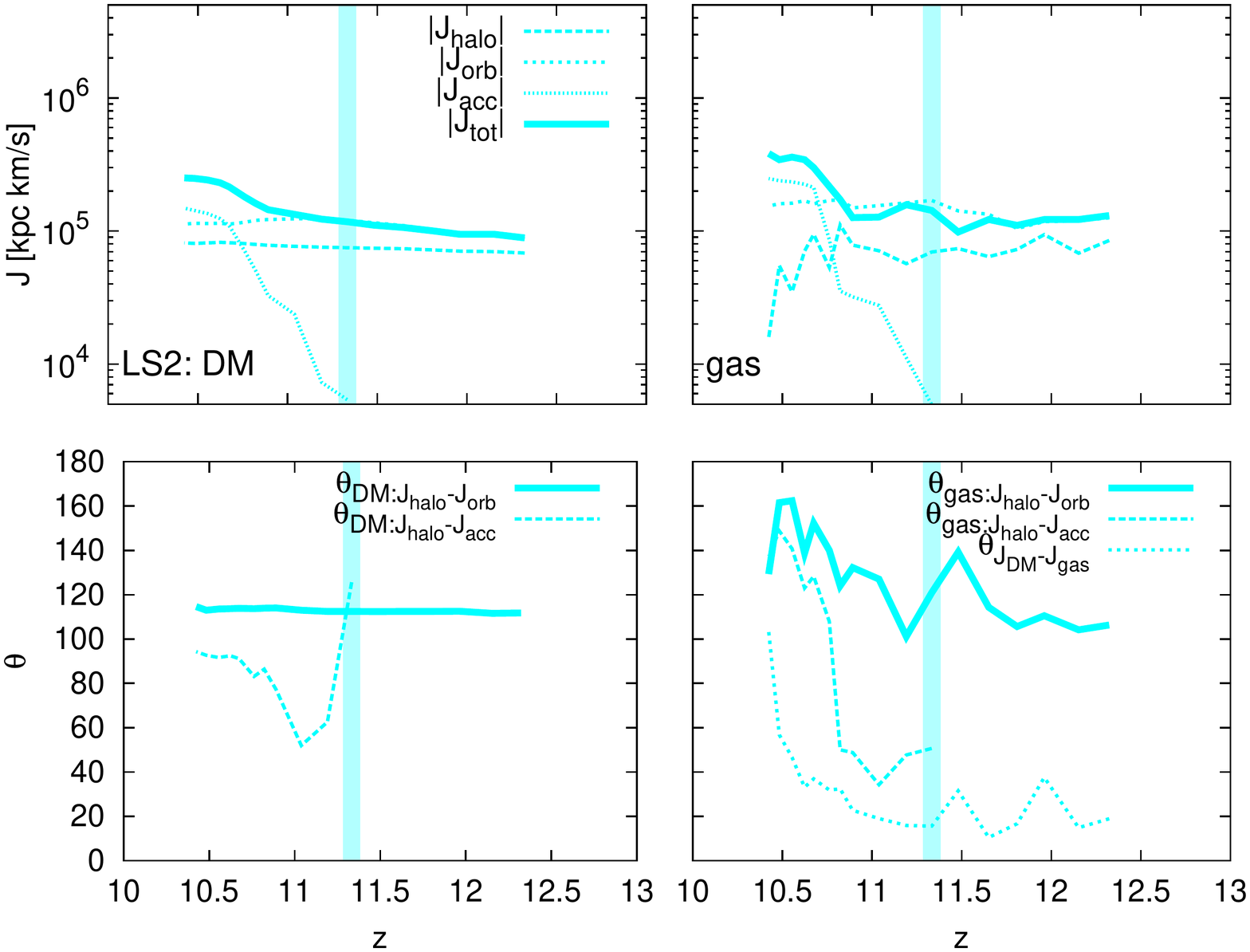}
\includegraphics[width=1.0\columnwidth]{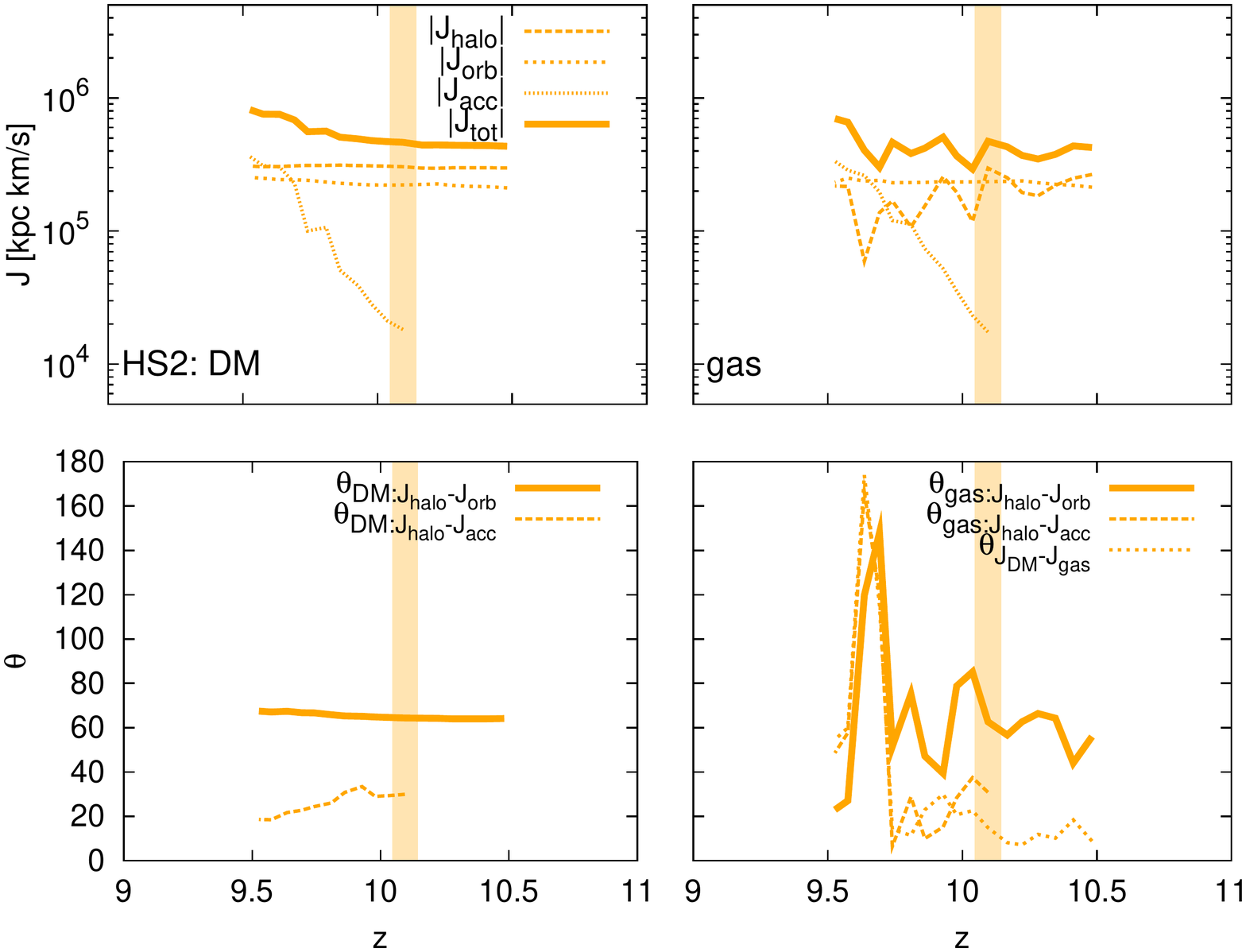}
\includegraphics[width=1.0\columnwidth]{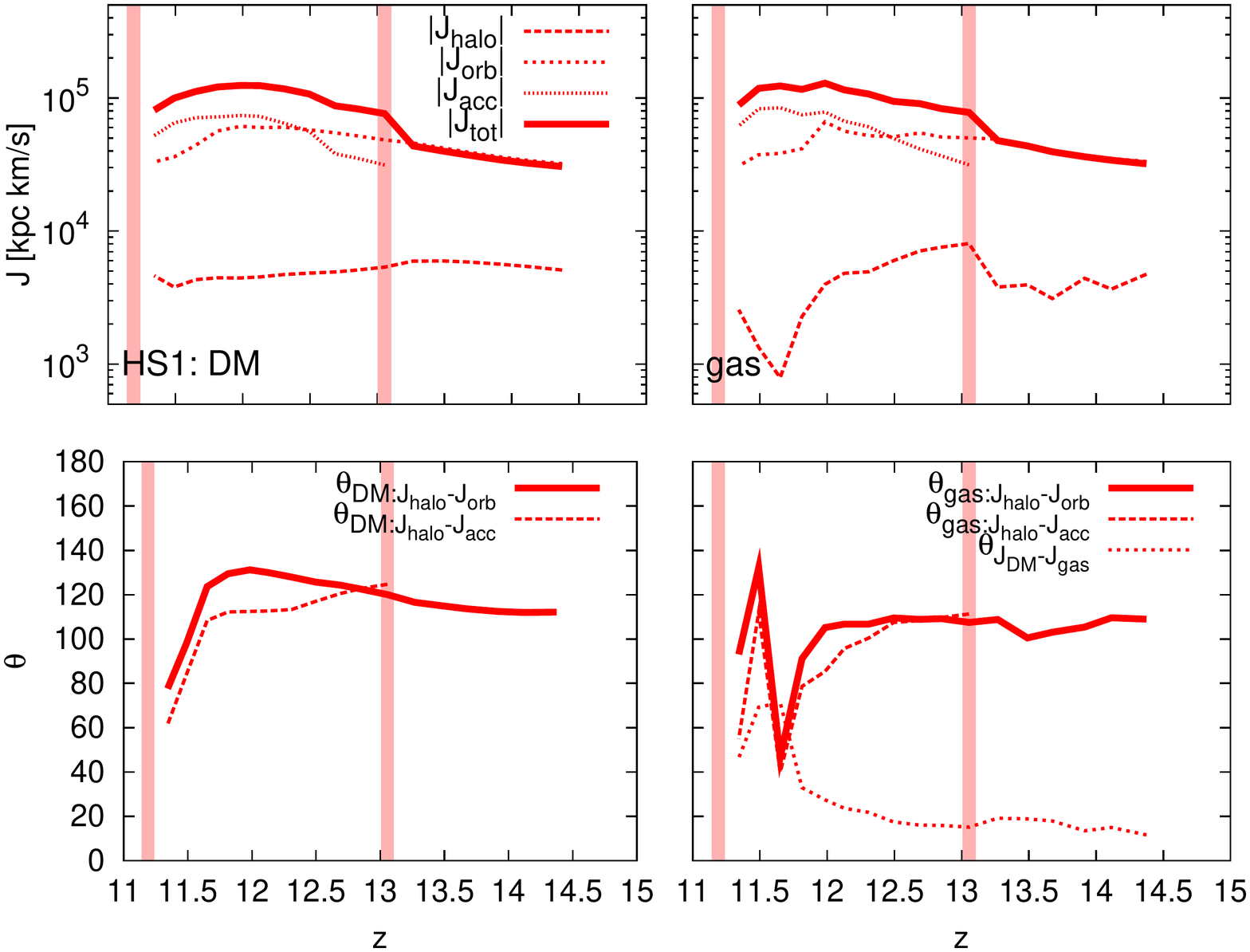}
\caption{Top panels: Specific angular momentum evolution in a merger event (DM at left, gas at right). The angular momentum 
is computed from the main progenitor's DM particles center of mass during its evolution with redshift. The long-dashed line shows the main halo
specific angular momentum; short-dashed line: the orbital angular momentum of the secondary progenitor of the final halo; dotted line: the specific angular momentum of the accreted material (which appears after the merger marked by the vertical line); solid line: the vectorial addition of these three components. Bottom panels: Angle between the three different
angular momentum vectors. The solid line is the angle
between the main halo's spin and the secondary progenitor's spin; long-dashed line: the angle between the main halo's
spin and the spin of the accreted material. The short-dashed line in the bottom right panel shows the gas-DM misalignment
angle of the particles associated to the main halo progenitor before the merger. The figure shows that the main halo spin
is much lower than the orbital angular momentum of the secondary progenitor before the merger (this is not a general trend,
see the HS2 case at bottom left). After the merger, the accreted material starts to increases its relative importance with respect 
to the orbital angular momentum. This continues until the accreted material dominates the source of angular momentum with respect 
to the main halo DM center of mass. The bottom panels show that different angular momentum components are not 
necessarily aligned due to the complex geometry of the process. This explains why the vectorial addition is not the dominant 
line in the top panels. The large fluctuations of the gas angles in the last stage of the evolutionary patch coincide with
the merger of the high density central regions of the two main progenitors.}
\label{fig:merger}
\end{figure*}

Keeping the above analysis in mind, it is useful to look at Fig.~\ref{fig:vectorfield} (and Fig.~\ref{fig:haloesmap}). This figure shows the 
tracer particles vector velocity field inside a sphere of radius R $\sim 300$ co-moving kpc around the corresponding halo colored 
by the initial SAM value at redshift $z\approx70$. The two red arrows located at the main halo center depict the DM SAM (solid line) 
and the gas SAM (dashed line) inside the virial radius. The vector field is shown at two different redshifts in order to illustrate the 
time evolution of the systems. The two most slowly rotating haloes, LS1 and LS2, show a mostly 
radial accretion channeled through a number of filaments with the vector spin placed at the convergence point of the filaments. On the other 
hand, the two fastest rotating systems, HS1 and HS2, develop a filament like structure with the halo vector spin more or less 
perpendicular to it. These suggest a relation between the structures around DM haloes, their gas in-falling pattern 
and their spin: haloes surrounded by a number of filaments tend to cancel out their accreted/merged SAM whereas haloes 
located in the middle of a filament tend to receive the accreted/merged 
SAM from a preferred direction resulting in a higher spin halo.

\begin{figure*}
\centering
\includegraphics[width=1.0\columnwidth]{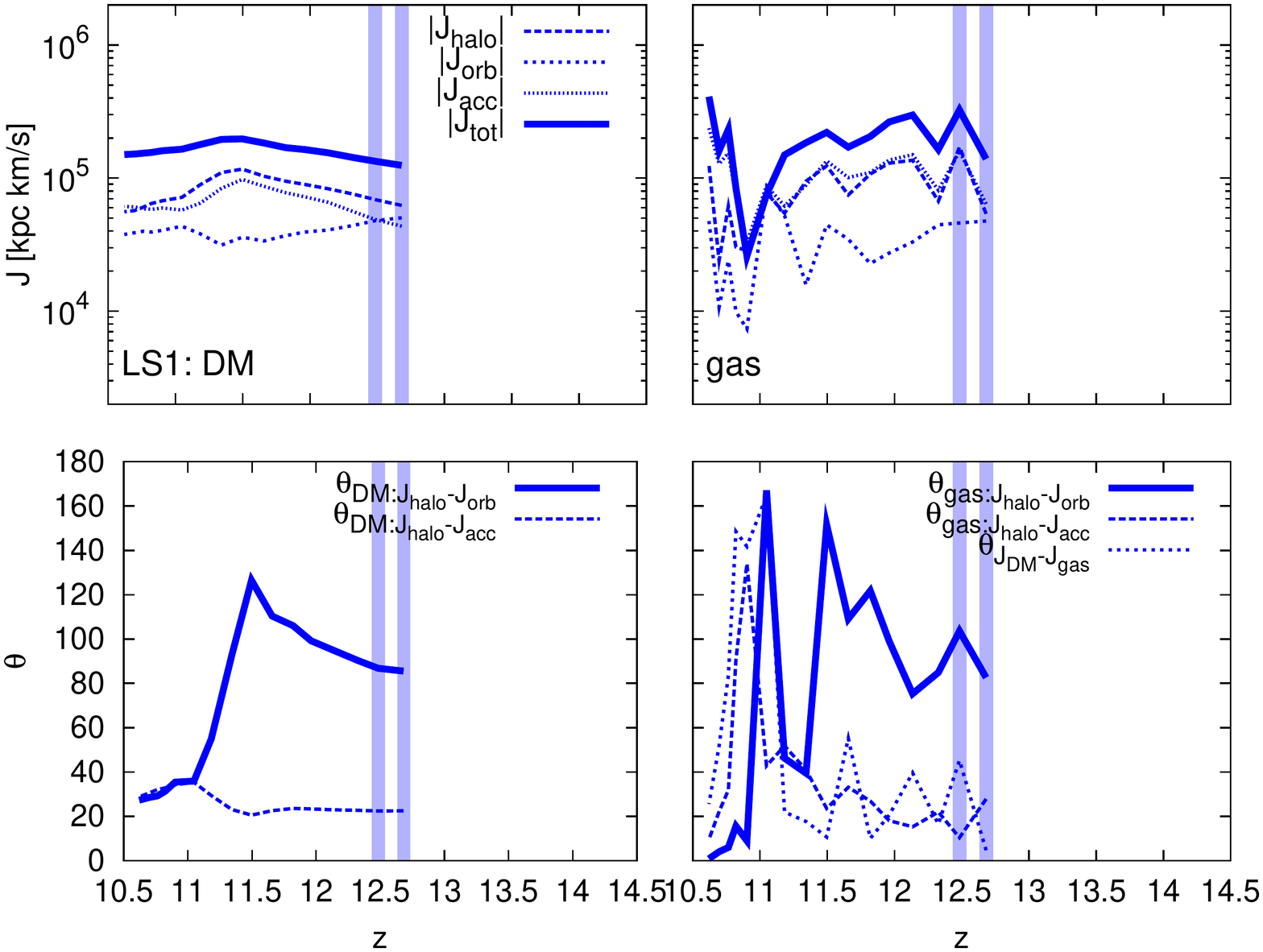}
\includegraphics[width=1.0\columnwidth]{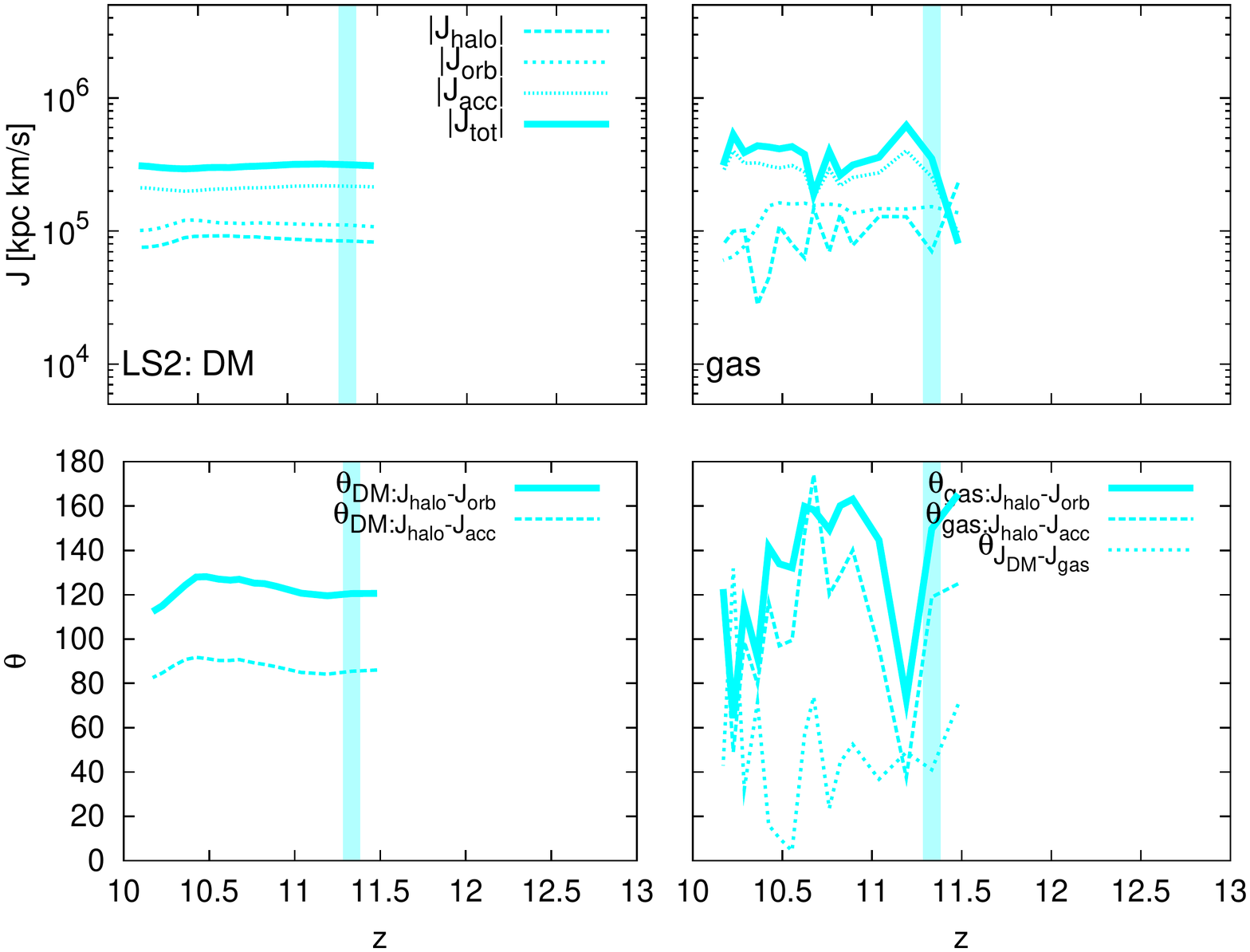}
\includegraphics[width=1.0\columnwidth]{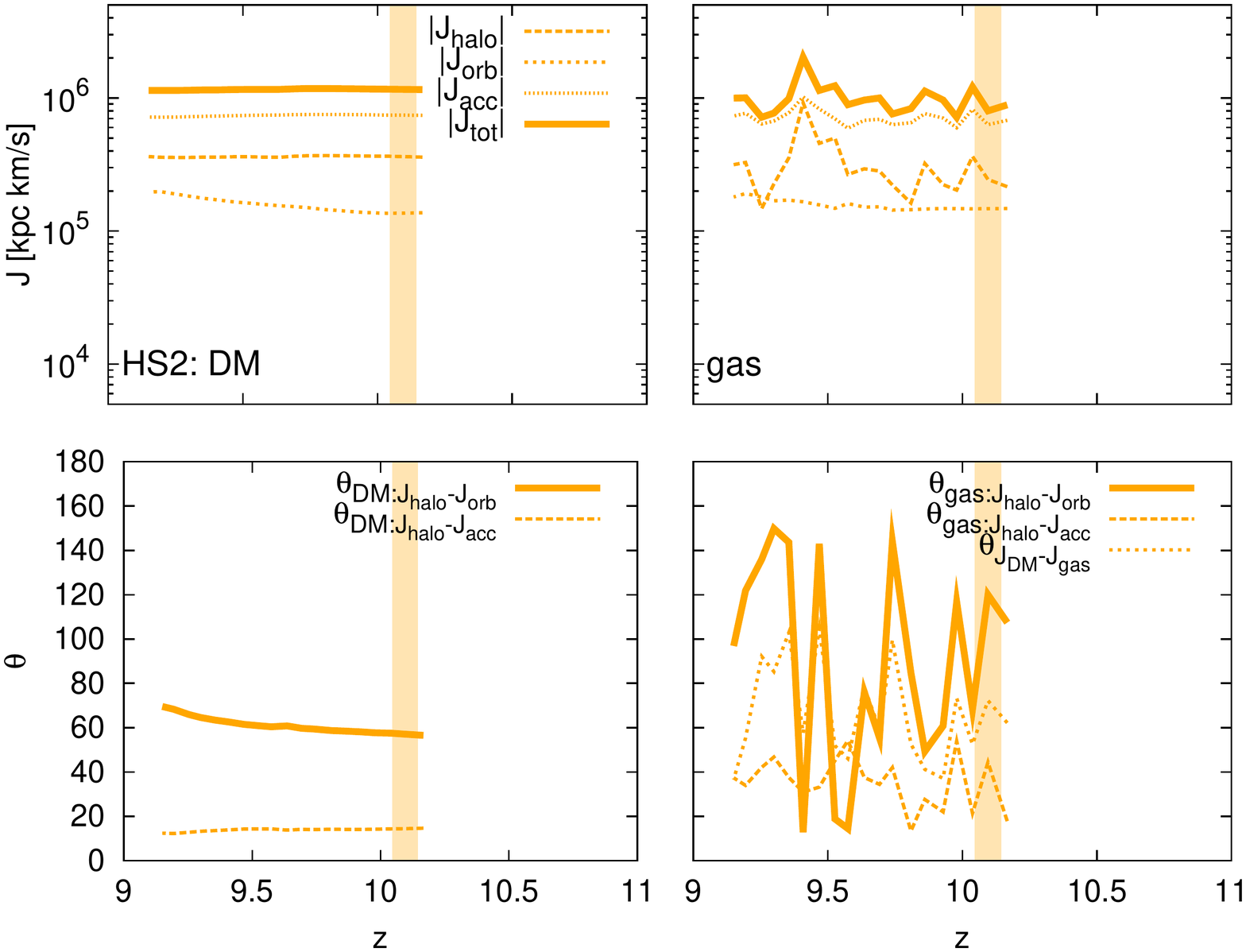}
\includegraphics[width=1.0\columnwidth]{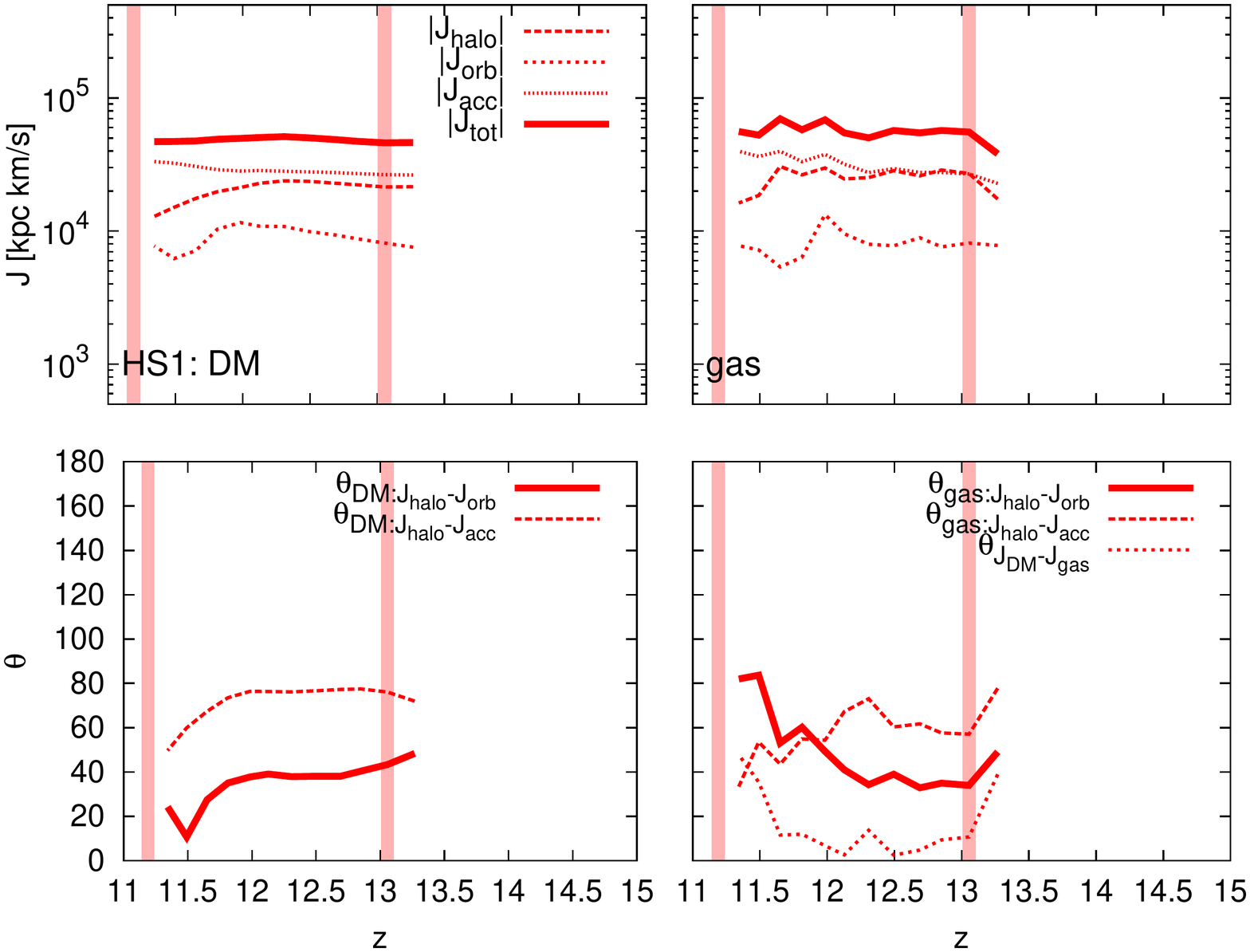}
\caption{Same as Fig.~\ref{fig:merger} but for the patch associated to particles (gas and DM) inside the corresponding
virial radius after the merger. The patch is selected at $\Delta z\approx1.5$ after the merger event marked by the vertical lines. }
\label{fig:mergerpatch}
\end{figure*}

We now turn our attention to the number of mergers that take place in each of the main haloes. Because the number of 
mergers happening in the main four haloes is $\approx 25$, they are too many as to describe each of them in detail. 
Therefore we will focus on a particular merger event per system\footnote{We will show that the merger events are a generic
process that can be described taking into account three SAM sources: the main halo one, the secondary halo one and the 
soft accreted one. With these three SAM sources we can fully describe the process.}. 

Fig.~\ref{fig:merger} shows the SAM and misalignment angle through a given merger event for each one of our haloes. As in 
previous figures the vertical lines shown the redshift of the merger event. The SAM evolution shown on the top rows 
is decomposed as follows: the long dashed line corresponds to the main halo SAM with respect to its own center of mass 
$|{\rm \vec{J}_{halo}}|$; 
the short dashed line shows the orbital SAM (with respect to the main halo center of mass) associated to the secondary halo going to 
merge with the main one, $|{\rm \vec{J}_{orb}}|$; the dotted line corresponds to the accreted SAM (with respect to the main halo 
center of mass), $|{\rm \vec{J}_{acc}}|$, note that this line will appear after the vertical line showing the merger 
event; finally the solid line is the modulus of the vectorial addition of the previous quantities: 
$|{\rm \vec{J}_{tot}}|=|{\rm \vec{J}_{halo}+\vec{J}_{orb}+\vec{J}_{acc}}|$.

In three of the four systems (LS1, LS2, HS1) $|{\rm \vec{J}_{orb}}|$ dominates the SAM before the merger. In the merger associated
to the HS2 halo, $|{\rm \vec{J}_{halo}}|$ is slightly larger than $|{\rm \vec{J}_{orb}}|$ before the merger. This results tell us
that from the main halo position the secondary halo going to merge brings a substantial amount of SAM to the new system (mainly 
due to the long radius ${\rm \vec{r}}$ in the SAM ${\rm \vec{r}\times\vec{v}}$ computation). Despite 
of this trend, it is interesting to note that after the merger the accreted material becomes the main source of SAM in the four 
mergers shown here. 

The bottom row of our four panels shows the misalignment angle between different components of the ${\rm \vec{J}_{tot}}$ vector. 
The different lines are: misalignment angle $\theta$ between ${\rm \vec{J}_{halo}}$ and ${\rm \vec{J}_{orb}}$ in solid 
line; $\theta$ between ${\rm \vec{J}_{halo}}$ and ${\rm \vec{J}_{acc}}$ in long dashed line and $\theta$ between ${\rm \vec{J}_{gas}}$ 
and ${\rm \vec{J}_{DM}}$ in short dashed line. All different SAM components are not aligned. The misalignment 
between all the different components is a consequence of partial cancellation of the total SAM. This phenomenon can be seen in 
the fact that $|{\rm \vec{J}_{tot}}|\leq|{\rm \vec{J}_{halo}|+|\vec{J}_{orb}|+|\vec{J}_{acc}}|$ in all cases. The misalignment angle 
associated to the gas component fluctuates significantly after the merger events. It can be explained 
by the pressure waves created when the dense central regions of the progenitors interact. Such process is able to create pressure 
torques, as explained in the previous section, resulting in a misalignment between the gas SAM components. Such torques are not present
in the DM.

We have also computed the SAM associated to a patch after the merger. In order to do so we have identified
the particles inside the virial radius of the main halo at $\Delta z\approx1.5$ after the merger event. Furthermore, we identified the
particles belonging to the main and to the secondary halo before the merger. With these definitions we marked as the accreted material
all the particles belonging to the final halo and not belonging either to the main nor the secondary halo. This is shown 
in Fig.~\ref{fig:mergerpatch}. The main difference with Fig.~\ref{fig:merger} is that from the patch center of mass the orbital angular 
momentum associated to the secondary halo is not the dominant spin source of the system (mainly due to the decrease of the radius ${\rm \vec{r}}$ 
in the SAM ${\rm \vec{r}\times\vec{v}}$ computation). In three (LS2, HS2 and HS1) of the four systems the main source of spin is associated 
to accreted material. The bigger the $\Delta z$ the higher 
the contribution from accreted material, thus we can conclude that in a system suffering a few mergers the softly accreted material 
will dominate the SAM source. Only in one (LS1) system the dominant source of spin is coming from the main halo. In 
this case the difference between the accreted and the main halo contribution is a factor $<$ 2: they are 
almost the same. The total DM SAM of the patch is almost constant in the considered redshift range. 

Table \ref{table1} and table \ref{table2} show all the relevant information about the merger processes with respect to the main
halo (labeled with 1) center of mass for DM and gas, respectively. Two interesting remarks based on these tables are that the LS systems have an
average mass merger ratio (${\rm M_2/M_1}$) lower than the HS systems and that the average angular momentum ratio (${\rm J_2/J_1}$)
is higher for the HS systems compared to the LS ones.

\begin{figure*}
\centering
\includegraphics[width=2.0\columnwidth]{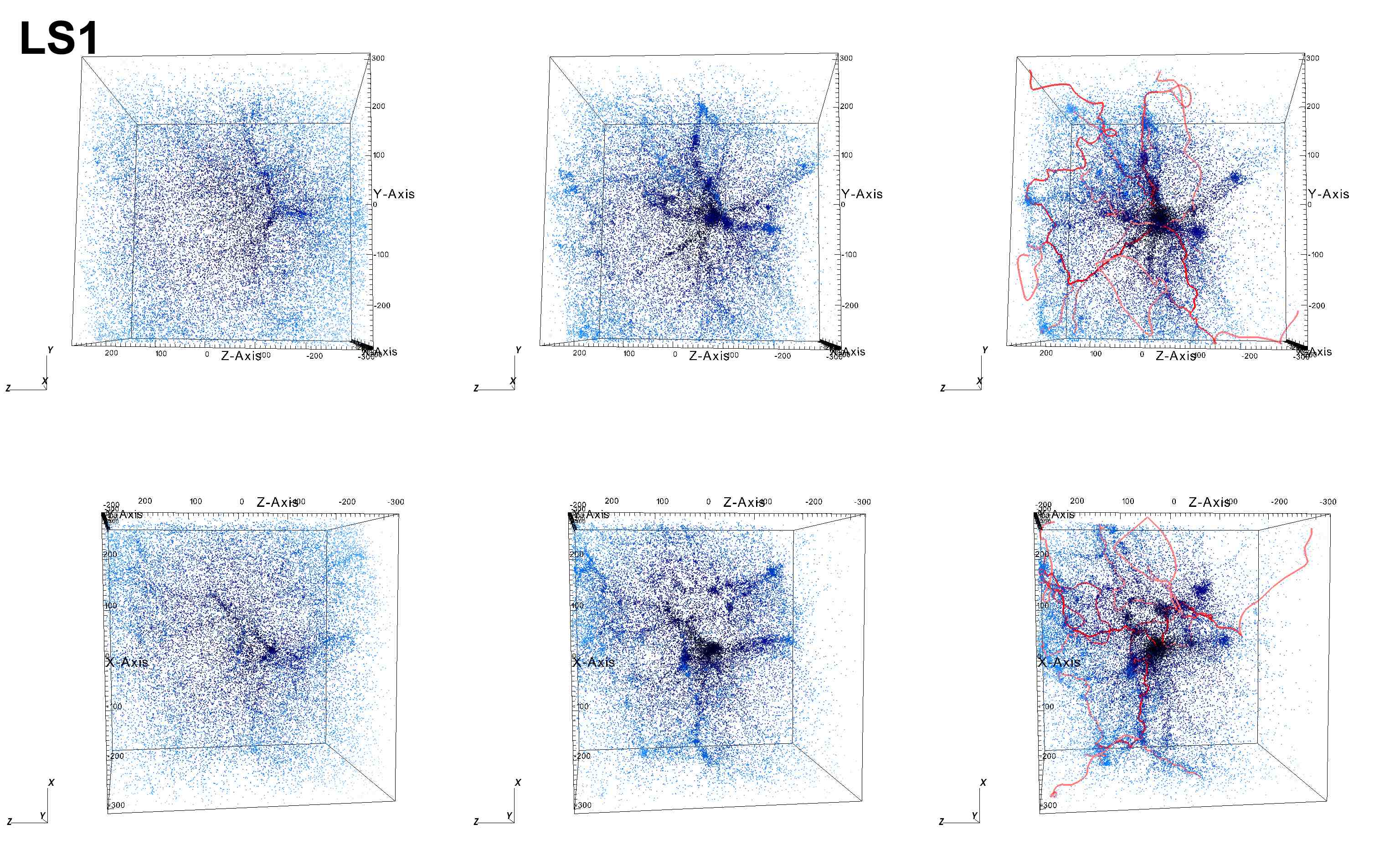}
\includegraphics[width=2.0\columnwidth]{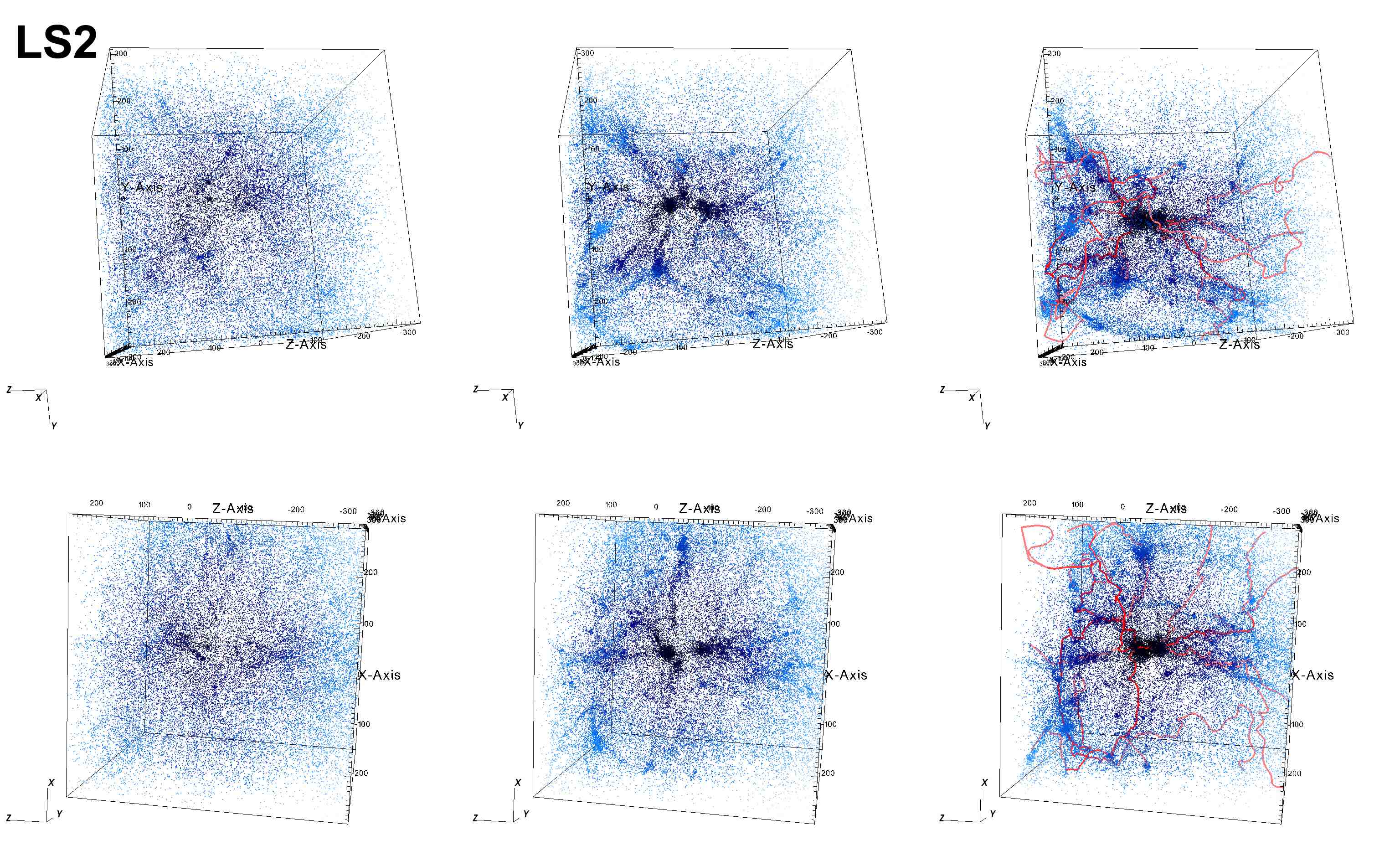}
\caption{LS DM evolution (in different columns) from different views (in different rows). The first two rows correspond to the LS1 
system and the last two rows correspond to the LS2 system. From the figure it is clear that both systems are located inside a knot
surrounded by a number of filaments marked by red lines in the last column, corresponding to the final redshift of the 
simulations, $z=9$. Such condition provides an efficient spin cancellation inside the virial radius.}
\label{fig:DMfilaments1}
\end{figure*}

\begin{figure*}
\centering
\includegraphics[width=2.0\columnwidth]{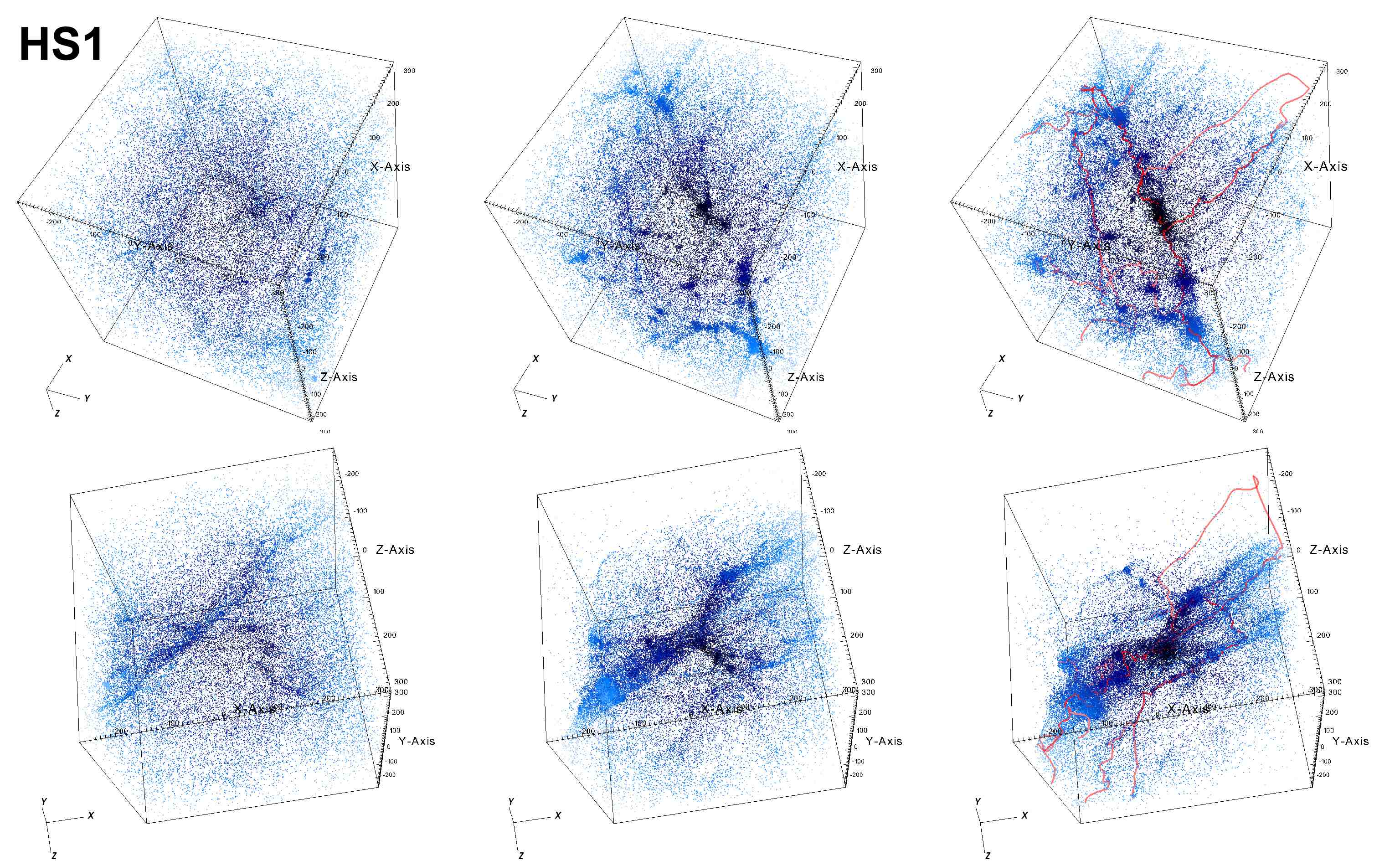}
\includegraphics[width=2.0\columnwidth]{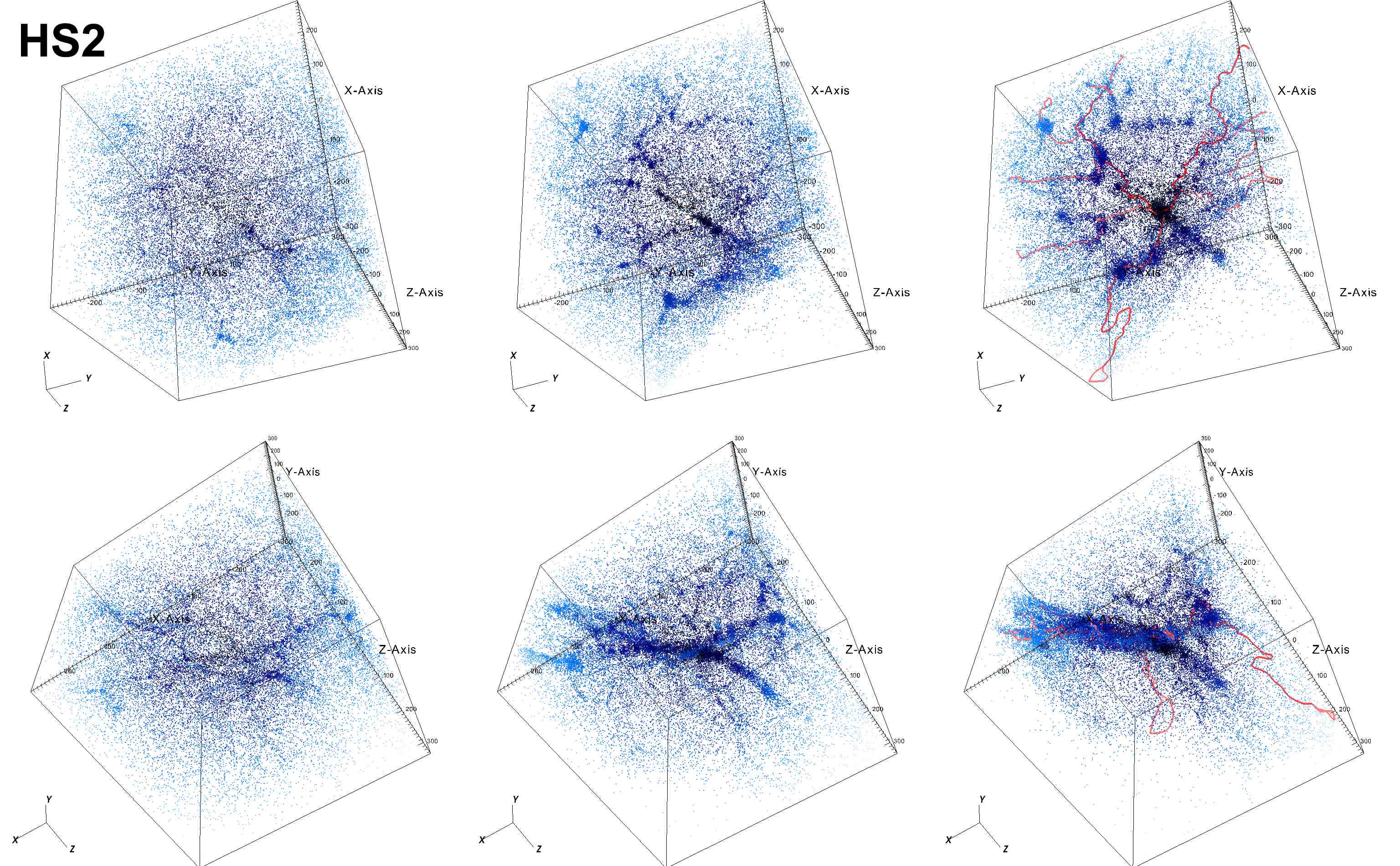}
\caption{Same as Fig.~\ref{fig:DMfilaments1} but for the HS systems. The first two rows correspond to the HS1 system and the last 
two rows correspond to the HS2 system. In this case the different views where selected to have a face-on view (first row)
and edge-on view (second row) of the collapsed DM wall. It is clear that the environmental/filamentary structure around the central haloes
is different compared to the LS system. The HS systems are located inside a filament belonging to a collapsed wall.}
\label{fig:DMfilaments2}
\end{figure*}

\subsection{Filaments and Environment.}

The environmental conditions around DM haloes seem to play a relevant role in the 
spin acquisition process (as previously suggested by \citet{Pichon+2011} but with a less detailed study as the one shown here). Because of this it is worth to study in more detail the filamentary structure of each halo and 
their position in the cosmic web. 

Here it is useful to come back to Fig.~\ref{fig:haloesmap}. If we look at the first row of the systems LS1 (top left)
and LS2 (top right) it is clear that around $z\approx11.5$ the main halos of these two systems were surrounded by a number 
(${\rm N_f\sim8}$) of filaments converging into a knot. Such special location creates a necessary condition for the SAM 
cancellation inside the virial radius. In such location 
all the material (accreted and merged) bringing a given amount of SAM will be able to cancel the total SAM value if from another direction is coming material with a comparable amount of SAM but pointing in another direction (which seems to be 
the general case as shown in Fig.~\ref{fig:merger} and Fig.~\ref{fig:flux}). This process can explain the low spin associated to 
these two haloes below $z\approx11$.

Looking at the first row of Fig.~\ref{fig:haloesmap} for the high spin systems, HS1 (bottom right) and HS2 (bottom left), we observe that 
at $z\approx11.5$ they are not in a knot of the cosmic web (as in the LS cases). While the HS2 halo is the composition of three
over-densities (the result of two mergers around $z\approx13$, see Fig.~\ref{fig:spinhalo}) inside a thin filament of length
${\rm L_f\ga200}$ co-moving kpc, the HS1 halo has a single over-density in the center and it is going to merge with a secondary 
halo belonging to a thin filament of length ${\rm L_f\approx100}$ co-moving kpc pointing to the top right corner of the box (this 
will be the merger associated to the vertical line at $z\approx10$ in Fig.~\ref{fig:spinhalo}). These
systems suffered important mergers at $z < 14$. Looking at table \ref{table1} the average merger mass ratio is higher for 
the HS systems compared with the LS systems and the same trend can be seen in the haloes spin ratio. Such features suggest that 
mergers are an important source of spin in the HS systems.   

In order to quantify the results presented above we compute the filamentary structure of our systems.
We compute at each snapshot the direction of the largest filaments connected to the main halo center.
Filaments are computed using the DisPerSE code \citep{disperse}, which computes for a particle distribution the 
Delaunay tessellation and uses discrete Morse theory to characterise the topology of the density field, i.e. critical/saddle 
points, lines and surfaces. Filaments are then traced by the saddle points connecting two local maximum critical points.
The advantages of this method are: i) it is scale-free thus filaments are purely defined by the point 
distribution, and a persistence threshold defined as the ``number of sigmas" with respect to Poisson noise.
ii) there is no smoothing involved in the density field computation; it uses directly the Delaunay tessellation, then it uses all 
available data specially useful to resolve filaments in poorly sampled fields with few particles. iii) only the direction of the 
filaments is needed, thus only filament skeletons are required and not the thickness or filament membership computed by more 
sophisticated methods.

The filaments directions are computed for each snapshot in the following way:
\begin{itemize}
\item We run DisPerSE code with a persistence threshold of $4\sigma$. This parameter is tuned to find only strong filaments which 
correlate very well with a visual inspection or other methods where typical filaments have a density contrast at least $\simeq 10$ 
times above average density \citep{Colberg+2005,GonzalezPadilla2010}.
\item From the filaments found in this way, we select only those which reach the central halo. That is, the filaments containing a critical point 
matching the FOF halo center.
\item Filaments are ranked by length and number of bifurcations/segments.
\item We compute the filament direction by averaging the filament path from the halo center up to a distance of $50$ kpc or 
$3 \times {\rm R_{vir}}$ depending on which value is smaller. We found in general that filaments are quite straight and radial 
at close distances, but beyond a few virial radii they warp and bifurcate into smaller filaments, making it hard to define 
a global filament direction.
\item The final most relevant filament is chosen by visual inspection among the few  filaments detected automatically, 
and results are in very good agreement between human and automated criterion, this last step is mostly intended to check consistency.
\end{itemize}

\begin{figure}
\centering
\includegraphics[width=1.0\columnwidth]{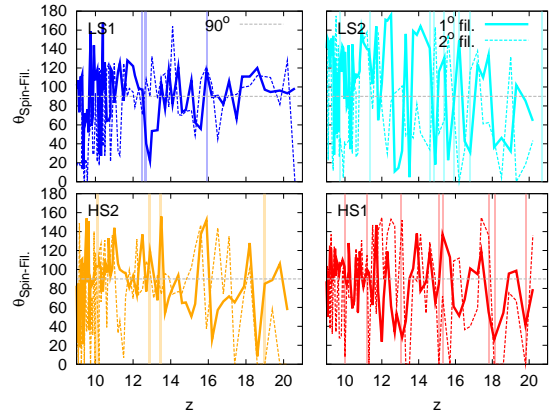}
\caption{Angle between the DM halo spin and the two main filaments associated to the same halo as a function of redshift 
for our four systems. The solid line 
corresponds to the main filament and the dashed line to the second main filament. The dashed gray line in the four 
panels marks a $90^\circ$ angle orientation.}
\label{fig:FilAngle}
\end{figure}

Fig.~\ref{fig:DMfilaments1} and \ref{fig:DMfilaments2} show the DM structure for our four systems with the filamentary structure
(in red lines) computed following the prescription described above. These figures show the 
DM evolution (in different columns) from two different almost perpendicular angles (in different rows). 

Fig.~\ref{fig:DMfilaments1} 
shows the evolution for the LS systems. LS1 corresponds to the first two rows and LS2 to the last two rows. It is clear that the 
LS systems are located in a knot inside the cosmic web, i.e. they are located at an over-dense region surrounded by a number of 
converging filaments (${\rm N_f\sim 8}$). Looking at the last column of the figure, it is worth to note that the filamentary inside-out extension does 
not vary too much from different view angles. Such conditions constrain the accreted material to converge on the knot from different directions providing an efficient SAM cancellation inside the virial radius. 

Fig.~\ref{fig:DMfilaments2} shows the HS systems. HS1 corresponds to the first two rows and HS2 to the last two rows. In this case
the two perpendicular views where selected to show a face-on view of the plane hosting the central halo (corresponding to the first 
row for each system) and an edge-on view of the same plane (corresponding to the second row for each systems). From the figure, it is possible to see how the collapse process of the wall hosting the central halo proceeds: it starts as a few parallel
filaments to finish as a thick filament (an angle view effect) crossing the box. From the face-on view of both planes it is clear that the central haloes 
are located in a filament belonging to the collapsed wall (marked by a red line crossing the box in the last column). For HS systems, such conditions tend to
constrain the accreted material to be mostly inside a filament belonging the wall. Such configuration makes the SAM cancellation process not as efficient as 
in the LS knot-located systems. This picture is supported by the fact that there are not too many red lines perpendicular to the DM plane associated
to the wall. This feature marks a clear difference with the LS systems where there is no a preferred plane hosting the filaments.

Fig.~\ref{fig:FilAngle} shows the angle between the DM SAM and the first two main filaments associated to the halo.
Our four systems show an angle fluctuating around $90^\circ$ for redshift $z\ga10$. It is interesting to note that mergers are correlated 
with sharp fluctuations of the angle. Such fluctuations are consequences of the re-configuration of the DM spin after 
violent merger events. Another interesting feature associated to mergers is the dispersion of the angle around $90^\circ$.
LS2 and HS1 show the larger dispersion and at the same time these two systems show the larger number of mergers which, as
mentioned before, can explain such angle deviations. The $\sim90^\circ$ orientation can be understood in the following way:
as shown before, most of the SAM is accreted through filaments onto the halo center, under such conditions 
any SAM component of the halo parallel to the filamentary SAM flux will be cancelled constraining the halo's SAM to a
plane perpendicular to the filamentary accretion. This spin orientation mechanism is similar to the one discussed by 
\citet{Codis+2012} in the high mass ($\ga10^{12}$M$_\odot$) range. 

\begin{figure}
\centering
\includegraphics[width=1.0\columnwidth,height=8cm]{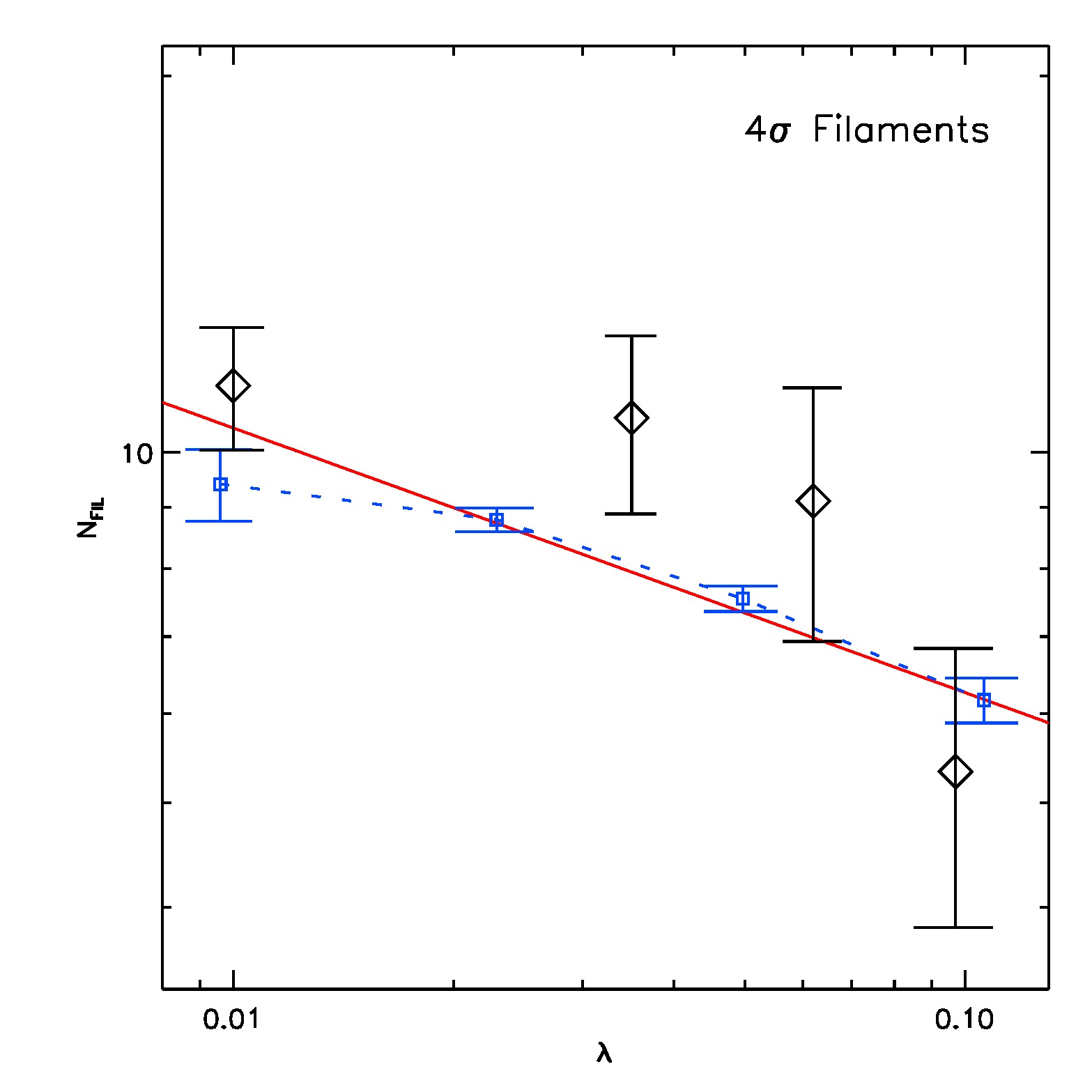}
\caption{Average spin parameter $\lambda$ as a function of the number of filaments ${\rm N_f}$ converging to the DM halo around
redshift $z=10$. Diamonds correspond to the four halos simulated in this paper including gas. The blue points are for 436 DM-only halos. 
There is a significant anti correlation between spin and number of filaments covering to the halo. The red solid line is the fit from eq.~\ref{eq:LvsN}.} 
\label{fig:LvsN}
\end{figure}

Fig~\ref{fig:LvsN} shows $\lambda$ against the number of filaments ${\rm N_f}$ around $z=10$ for the 4 halos simulated in this paper (diamond symbols). 
We found an anti-correlation between the number of filaments that converge at the DM halo and its spin value.

Despite of the small data sample (4 haloes), we can see a clear anti-correlation supporting our previous discussion.  Because the above anti-correlation is so 
interesting, we investigated it in DM-only  simulations; these consist of simulations with ${\rm N_p} = 512^3$, with a particle mass of $10^6$ M$_{\odot}$. The blue points 
of Fig.~\ref{fig:LvsN} show the dependence between the halo spin and the number of filaments connected to the halo center.
We show results for $436$ halos in the mass range
$5\times10^{8}-5\times10^{9}M_{\odot}$, selecting 4 spin bins with error bars showing the $1\sigma$ uncertainty from bootstrapping.
Filaments are computed using Disperse code with a $4\sigma$ detection threshold. There is a significant trend that confirms the one found with the 4 halos simulated in this paper: lower
spin halos have more filaments connected to them  while higher spin halos have less
filaments connected them. The correlation can be written as

\begin{equation}
\lambda \approx 0.05 \times \left(\frac{7.6}{\rm N_f}\right)^{5.1}
\label{eq:LvsN}
\end{equation}

\begin{figure}
\centering
\includegraphics[width=9.0cm,height=11.0cm]{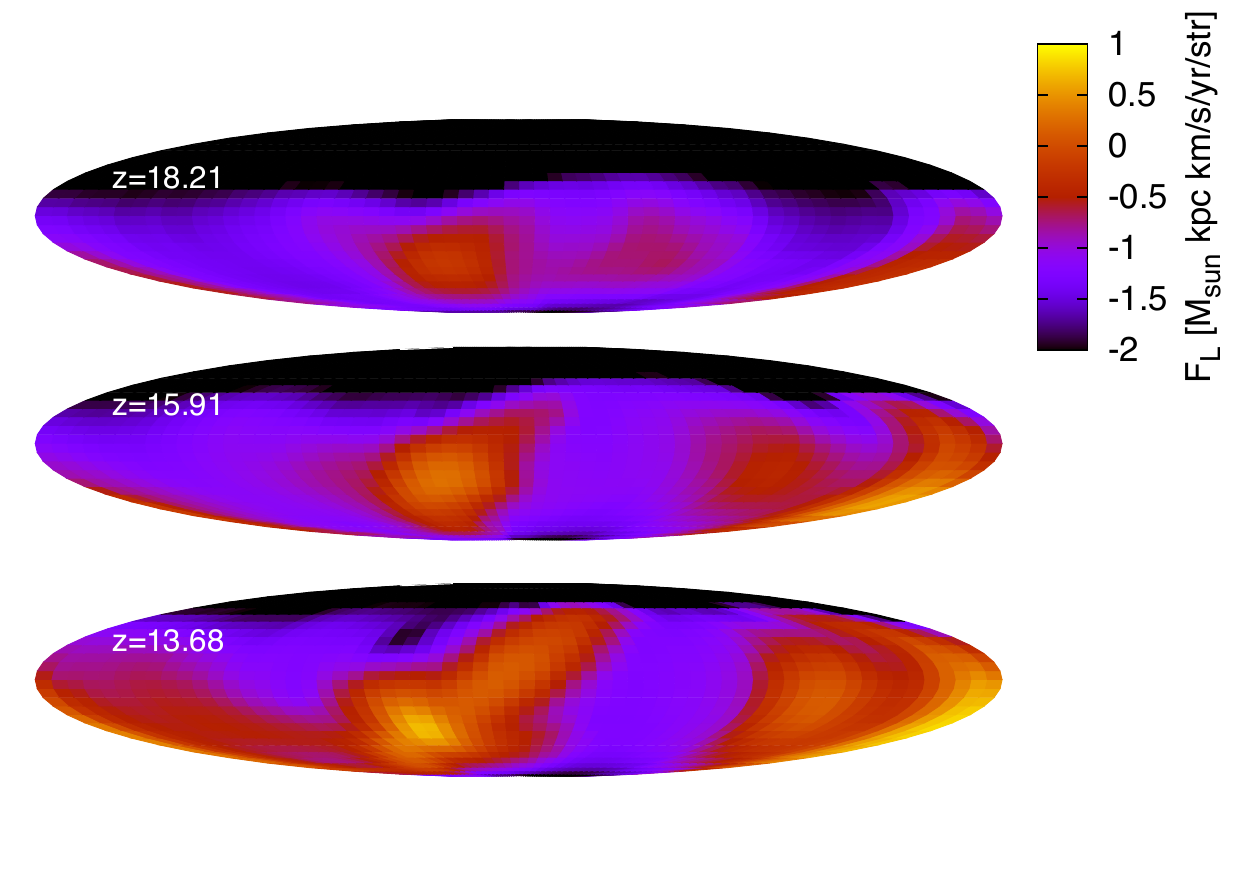}
\caption{Mass weighted SAM flux for the HS2 system at different redshifts. The SAM accretion evolves from a rather homogeneous flux at $z=18.21$ to a well defined filamentary accretion at $z=13.68$ (the filament is defined by the strong accretion (yellow blob) at angle $\approx \pi/2$. This transition from an homogeneous to filamentary accretion explains the spin parameter evolution of HS2 in Fig.~\ref{fig:lambda}.}
\label{fig:fluxFil}
\end{figure}

At this point it is  interesting to come back to Fig.~\ref{fig:lambda}. The $\lambda$ parameter
changes significantly with redshift for each halo. In particular it is interesting to understand $\lambda$ variations in both  LS1
and HS2 systems. For these systems the spin parameter changes from $\sim0.1$ to $\sim0.01$ during their evolution. Such changes can be fully understood
following the picture presented in the analysis above. The LS1 system starts with  $\lambda\approx0.03-0.04$ at $z\approx20-16$.
At this stage the main halo is inside a small filament. At $z\approx16$ the system receives a merger through the filament increasing its spin
parameter till $\lambda\ga0.1$. After that the system starts to accrete mass isotropically till it forms a clear convergent filamentary 
structure around it resulting in a low spin system at $z\sim9$. The HS2 system starts with a high spin parameter at $z\approx20-19$.
After the merger at $z\approx19$ the system does not belong to a filament and starts to accrete mass rather isotropically. This explains the
reduction of the spin till $\sim0.01$ at $z\approx16$. At this stage the system has developed a clear filamentary
structure and most of the accreted material starts to come from this preferential direction. The system continues
accreting mass till the it receives two mergers at $z\ga13$ when $\lambda$ reaches the value $\sim0.1$; see Fig.~\ref{fig:fluxFil}.

The emerging picture from the above simulations seems to be one in which the merging process, which in turn is determined 
by environmental location, determines the final value of the halo spin. 

\section{Conclusions}
\label{conclusions}

We have performed numerical simulations of atomic cooling halos with the aim of understanding the origin of angular momentum in 
galaxies. Atomic cooling halos maybe rare objects in the sky, but they are pristine sites of galaxies formation where the 
complications of star formation and subsequent feedback are avoided. They therefore provide an excellent testbed to explore 
our theories for galaxy formation. 

The numerical experiments we have performed are for four values of the halo spin parameter $\lambda = 0.01, 0.04, 0.06, 0.1$ thus 
sampling the 1 and 2$\sigma$ ranges of the theoretically predicted LCDM probability distribution function for dark matter halos. 
We have shown that before 
turn-around the spin value of the halos follows that predicted by tidal torque theory and that after turn-around the spin value 
is influenced by the number and direction of mergers and accreting dark matter. While this might lead one to think that after turn 
around the angular momentum of the gas and dark matter will be totally random, we show this not to be the case. 

The final angular momentum of dark matter is strongly correlated to the morphology of the cosmic web where the halos 
reside. In this respect, halos with low angular momentum are those residing at the center of two or more filament crossings, 
while halos with high angular momentum are those residing in the middle of a filament and away from the knots of the cosmic web. 
There is a clear anti-correlation between the number of filaments and the final angular momentum. The physical picture that we 
find is that where many filaments cross, the flow of dark matter is from many directions and thus, by randomization, one ends 
up destroying angular momentum as it comes from different directions. On the other hand, if the halo is positioned in the middle 
of a filament, there is only one preferred direction from where dark and baryonic matter is accreted thus preserving the angular 
momentum originally determined by tidal torque theory. On the other hand, we find that the angular momentum of the gas is dominated by gradient pressures and that this causes fluctuations and misalignment with respect to the DM angular momentum, thus making it more unpredictable.

Our findings point to a picture in which the angular momentum of DM halos is obtained by features in the LCDM that are very deterministic: 
the rate at which other halos and dark matter are being accreted, and the morphology of the cosmic web. However, the angular momentum of the gas is driven by the gas physics itself, making it disconnected to the one of the underlying DM.

\section*{Acknowledgments}

ZH acknowledges support from NASA grant NNX11AE05G. 
JP and RJ acknowledge support from Mineco grant FPA2011-29678-C02-02.
JP acknowledges support from Anillo de Ciencia y Tecnologia (Project ACT1101). JP also thanks Yohan Dubois for  
providing the tracer particle patch for the RAMSES code. REG was supported by Proyecto Financiamiento Basal PFB-06 'Centro de Astronomia y Tecnologias Afines' and Proyecto Comit\'e Mixto ESO 3312-013-82. The Geryon cluster at the Centro de Astro-Ingenieria UC was extensively used for the analysis calculations performed in this paper. The Anillo ACT-86, FONDEQUIP AIC-57, and QUIMAL 130008 provided funding for several improvements to the Geryon cluster.

\section*{Appendix}

The Tables in this section list merger related parameters for the four halos studied in this paper.
Label 1 denotes  quantities associated to the main halo, while label 2  the quantities associated 
to the secondary progenitor, the label orb denotes quantities associated to the orbital angular momentum 
associated to the secondary progenitor and the label acc quantities associated to the accreted 
angular momentum.

\begin{table*}
\caption{DM Merger Events.}
\centering
\begin{tabular}{l l c c c c c r r r}
\hline\hline
     &       &       &         &       &         &      &       &       &       \\
Sim. & Event &  z    & M$_1$           &  $\frac{{\rm M}_2}{{\rm M}_1}$ & J$_1$ & $\frac{{\rm J}_2}{{\rm J}_1}$ & $\theta_{\vec{{\rm J}}_1,\vec{{\rm J}}_2}$  &  $\theta_{\vec{{\rm J}}_1,\vec{{\rm J}}_{\rm orb}}$ & $\theta_{\vec{{\rm J}}_1,\vec{{\rm J}}_{\rm acc}}$ \\
Name &       &       & [M$_{\odot}$]   &         & [kpc km/s]&  &       &       \\
\hline\hline
LS1  &       &       &         &       &         &      &       &       &       \\
     & Merg1 & 12.68 & 3.89(8) & 0.34  & 3.26(4) & 0.40 & 137.3 & 104.6 &  64.0 \\
     & Merg2 & 12.86 & 3.46(8) & 0.20  & 4.04(4) & 0.01 &  55.5 & 134.9 &  24.1 \\
     & Merg3 & 16.17 & 1.03(8) & 0.12  & 3.38(3) & 0.11 &  67.5 &  51.5 &  36.9 \\
     & Merg4 & 26.75 & 2.98(6) & 0.42  & 7.65(0) & 0.19 & 112.7 & 157.8 & 139.0 \\
Avg. &       &       &         & 0.27  &         & 0.18 &  93.3 & 112.2 &  66.0 \\
     &       &       &         &       &         &      &       &       &       \\
LS2  &       &       &         &       &         &      &       &       &       \\
     & Merg1 &  9.80 & 1.48(9) & 0.72  & 3.93(5) & 0.15 &  71.9 & 123.6 &  87.9 \\
     & Merg2 & 11.48 & 7.77(8) & 0.17  & 7.43(4) & 0.15 & 147.6 & 112.3 &  26.6 \\
     & Merg3 & 14.85 & 1.68(8) & 0.12  & 1.29(4) & 0.03 &  33.0 &  73.4 &  15.5 \\
     & Merg4 & 15.11 & 1.24(8) & 0.31  & 8.57(3) & 0.04 &  73.5 &  55.6 &  50.0 \\
     & Merg5 & 15.60 & 9.64(7) & 0.14  & 4.46(3) & 0.02 &  21.4 &  61.3 &  76.8 \\
     & Merg6 & 16.19 & 7.09(7) & 0.31  & 3.63(3) & 0.16 & 119.5 & 122.2 &  22.9 \\
     & Merg7 & 16.52 & 6.18(7) & 0.17  & 3.26(3) & 0.07 &  50.9 &  95.2 &   9.7 \\
     & Merg8 & 17.16 & 4.30(7) & 0.20  & 7.65(2) & 0.08 &  93.2 &  75.3 &  61.5 \\
     & Merg9 & 21.13 & 3.74(6) & 0.67  & 1.14(1) & 2.57 & 114.1 & 126.9 &  45.9 \\
Avg. &       &       &         & 0.31  &         & 0.36 &  80.5 &  94.0 &  44.1 \\
     &       &       &         &       &         &      &       &       &       \\
HS2  &       &       &         &       &         &      &       &       &       \\
     & Merg1 & 10.16 & 9.73(8) & 0.12  & 2.97(5) & 0.02 &  26.7 &  64.2 &   4.7 \\
     & Merg2 & 13.06 & 1.82(8) & 0.89  & 1.42(4) & 0.83 & 126.1 &  36.6 &  20.8 \\
     & Merg3 & 13.67 & 1.42(8) & 0.38  & 7.99(3) & 0.12 & 115.1 &  19.7 &   5.6 \\
     & Merg4 & 19.36 & 4.08(6) & 1.07  & 1.98(1) & 1.85 &  74.9 &  86.3 &  76.0 \\
Avg. &       &       &         & 0.62  &         & 0.71 &  85.7 &  51.7 &  26.8 \\
     &       &       &         &       &         &      &       &       &       \\
HS1  &       &       &         &       &         &      &       &       &       \\
     & Merg1 & 10.04 & 8.21(8) & 0.48  & 1.84(5) & 0.20 & 105.1 &  27.2 &  18.7 \\
     & Merg2 & 11.34 & 4.31(8) & 0.29  & 3.61(4) & 0.26 & 154.4 & 152.0 &  52.9 \\
     & Merg3 & 13.26 & 9.13(7) & 0.64  & 5.93(3) & 0.67 &  23.6 & 116.6 & 119.4 \\
     & Merg4 & 15.33 & 3.10(7) & 0.49  & 5.93(2) & 0.78 & 132.8 & 129.2 & 143.6 \\
     & Merg5 & 15.65 & 2.72(7) & 0.22  & 7.07(2) & 0.80 & 101.2 & 175.8 &  14.9 \\
     & Merg6 & 18.13 & 7.57(6) & 0.61  & 8.48(1) & 0.38 & 129.3 & 114.3 & 130.8 \\
     & Merg7 & 18.56 & 5.78(6) & 0.26  & 5.54(1) & 0.11 & 129.4 & 134.9 &  13.9 \\
     & Merg8 & 20.22 & 2.26(6) & 0.52  & 1.02(1) & 0.23 & 123.1 & 104.4 &  59.1 \\
Avg. &       &       &         & 0.44  &         & 0.43 & 112.4 & 119.3 &  69.2 \\		
     &       &       &         &       &         &      &       &       &       \\
\hline\hline
\end{tabular}
\label{table1}
\end{table*}

\begin{table*}
\caption{Gas Merger Events.}
\centering
\begin{tabular}{l l c c c c c r r r}
\hline\hline
     &       &       &         &       &         &      &       &       &       \\
Sim. & Event &  z    & M$_1$           &  $\frac{{\rm M}_2}{{\rm M}_1}$ & J$_1$ & $\frac{{\rm J}_2}{{\rm J}_1}$ & $\theta_{\vec{{\rm J}}_1,\vec{{\rm J}}_2}$  &  $\theta_{\vec{{\rm J}}_1,\vec{{\rm J}}_{\rm orb}}$ & $\theta_{\vec{{\rm J}}_1,\vec{{\rm J}}_{\rm acc}}$ \\
Name &       &       & [M$_{\odot}$]   &         & [kpc km/s]&  &       &       \\
\hline\hline
LS1  &       &       &         &       &         &      &       &       &       \\
     & Merg1 & 12.68 & 6.31(7) & 0.36  & 2.41(4) & 0.54 & 135.5 & 100.9 &  68.1 \\
     & Merg2 & 12.86 & 5.62(7) & 0.26  & 2.46(4) & 0.07 &  28.9 & 143.8 &   5.5 \\
     & Merg3 & 16.17 & 1.64(7) & 0.11  & 2.56(3) & 0.10 &  43.7 &  61.8 &  38.3 \\
     & Merg4 & 26.75 & 2.82(5) & 0.24  & 1.35(0) & 0.42 &  78.5 & 125.2 &  96.9 \\
Avg. &       &       &         & 0.24  &         & 0.28 &  71.7 & 107.9 &  52.2 \\
     &       &       &         &       &         &      &       &       &       \\		
LS2  &       &       &         &       &         &      &       &       &       \\
     & Merg1 &  9.80 & 2.56(8) & 0.73  & 5.47(5) & 0.39 &   5.3 &  48.4 &   3.9 \\
     & Merg2 & 11.48 & 1.28(8) & 0.22  & 7.37(4) & 0.28 & 156.1 & 139.1 &   9.7 \\
     & Merg3 & 14.85 & 2.54(7) & 0.12  & 1.13(4) & 0.04 &  54.2 &  76.6 &  23.2 \\
     & Merg4 & 15.11 & 2.19(7) & 0.30  & 1.31(4) & 0.02 &  98.4 &  73.0 &  25.4 \\
     & Merg5 & 15.60 & 1.58(7) & 0.13  & 5.42(3) & 0.02 &  23.6 &  25.1 &  44.6 \\
     & Merg6 & 16.19 & 1.13(7) & 0.29  & 4.27(3) & 0.07 & 106.4 & 143.3 &   9.1 \\
     & Merg7 & 16.52 & 9.46(6) & 0.21  & 2.66(3) & 0.07 & 131.6 & 136.5 &  11.3 \\
     & Merg8 & 17.16 & 6.57(6) & 0.16  & 8.94(2) & 0.02 &  59.6 &  74.1 &  54.2 \\
     & Merg9 & 21.13 & 4.92(5) & 0.59  & 1.79(0) & 0.83 &  82.4 &  64.8 &  73.4 \\
Avg. &       &       &         & 0.31  &         & 0.19 &  79.7 &  86.8 &  28.3 \\
     &       &       &         &       &         &      &       &       &       \\		
HS2  &       &       &         &       &         &      &       &       &       \\
     & Merg1 & 10.16 & 1.65(8) & 0.11  & 2.51(5) & 0.05 &  25.8 &  56.6 &   4.1 \\
     & Merg2 & 13.06 & 2.85(7) & 0.95  & 9.13(3) & 1.35 & 115.0 &  71.9 &  60.1 \\
     & Merg3 & 13.67 & 2.22(7) & 0.43  & 5.54(3) & 0.32 &  90.1 &  44.0 &  12.7 \\
     & Merg4 & 19.36 & 5.60(5) & 0.73  & 1.10(1) & 0.82 &  92.5 &  86.5 &  64.4 \\
Avg. &       &       &         & 0.56  &         & 0.64 &  80.9 &  64.8 &  35.3 \\
     &       &       &         &       &         &      &       &       &       \\		
HS1  &       &       &         &       &         &      &       &       &       \\
     & Merg1 & 10.04 & 1.41(8) & 0.57  & 2.41(5) & 0.17 &  91.2 &  31.3 &   9.2 \\
     & Merg2 & 11.34 & 7.02(7) & 0.31  & 4.18(4) & 0.30 & 166.2 & 155.9 &  18.9 \\
     & Merg3 & 13.26 & 1.54(7) & 0.63  & 3.79(3) & 1.09 &  43.3 & 108.7 & 110.9 \\
     & Merg4 & 15.33 & 4.52(6) & 0.48  & 6.66(2) & 0.38 &  89.0 & 115.2 & 129.7 \\
     & Merg5 & 15.65 & 3.92(6) & 0.22  & 6.78(2) & 0.07 & 112.4 & 165.5 &  15.2 \\
     & Merg6 & 18.13 & 1.01(6) & 0.54  & 3.03(1) & 0.48 &  45.1 &  55.1 &  67.4 \\
     & Merg7 & 18.56 & 6.86(5) & 0.21  & 1.74(1) & 0.07 &  20.1 &  85.3 &   5.7 \\
     & Merg8 & 20.22 & 2.04(5) & 0.41  & 2.49(0) & 0.35 & 146.6 &  66.6 &  51.2 \\
Avg. &       &       &         & 0.42  &         & 0.36 &  89.2 &  98.0 &  51.0 \\		
     &       &       &         &       &         &      &       &       &       \\
\hline\hline
\end{tabular}
\label{table2}
\end{table*}

\end{document}